\documentclass[a4paper,fleqn]{cas-dc}

\usepackage[numbers,sort&compress]{natbib}

\def\tsc#1{\csdef{#1}{\textsc{\lowercase{#1}}\xspace}}
\tsc{WGM}
\tsc{QE}

\usepackage{tabularx}
\usepackage{array}
\usepackage{booktabs}

\begin{document}
\let\WriteBookmarks\relax
\def\floatpagepagefraction{1}
\def\textpagefraction{.001}

\shorttitle{Simulation of Non-Markovian Quantum Accelerated Dynamics via Time-Fractional Schr\"odinger Equation}    

\shortauthors{Wei et~al.}  

\title [mode = title]{Simulation of Non-Markovian Quantum Accelerated Dynamics via Time-Fractional Schr\"odinger Equation}  

\tnotemark[1]

\tnotetext[1]{This work is supported by Basic Research Program Youth Project of Lianyungang, China (Grant No. JCYJ2523), and Haizhou Bay Talent Innovation Program of Jiangsu Ocean University (Grant No. KQ25062).}

%

\author[1]{Dongmei Wei}[orcid=0009-0009-0445-5321]



\credit{Conceptualization, Methodology, Software, Formal analysis, Investigation, Supervision, Writing - original draft}

\author[1]{Junxiang Wang}
\credit{Software, Formal analysis, Writing - review \& editing}
\author[1]{Hanxiu Xu}
\credit{Validation, Data curation, Writing - review \& editing}
\author[1]{Cancan Chen}
\credit{Methodology, Validation, Investigation, Supervision, Writing - review \& editing}

\author[1]{Jiaying Wu}[orcid=0000-0001-5184-4295]
\cormark[1]
\ead{jiaying@jou.edu.cn}
\credit{Supervision, Project administration, Validation, Writing - review \& editing}

\affiliation[1]{organization={School of Computer Engineering, Jiangsu Ocean University},
            addressline={59 Cangwu Road, Haizhou District}, 
            city={Lianyungang},
            postcode={222005}, 
            state={Jiangsu},
            country={China}}

\cortext[cor1]{Corresponding author}


\begin{abstract}
The Time-Fractional Schr\"odinger Equation (TFSE) is an effective tool for simulating the dynamics of non-Markovian quantum systems. The Quantum Speed Limit (QSL) time characterizes the minimum time required for the evolution of a non-Markovian quantum system. In this paper, Wei's TFSE is employed to simulate the non-Markovian quantum accelerated evolution process in the Resonant Dissipative Jaynes-Cummings (RDJC) model. By solving the QSL time of a time-fractional single-qubit open system, the enhancement mechanism of the system evolution speed induced by the non-Markovian memory effects of the environment is revealed. Further studies show that the optimized acceleration of the system evolution can be achieved by jointly regulating the fractional order, coupling strength, and photon number. Comparative analyses indicate that Wei's TFSE can accurately capture the non-Markovian accelerated dynamical features of the system over the entire fractional order range, whereas Naber's TFSE is applicable only within a limited fractional order interval. In addition, the comparisons of the average simulation time for calculating the dynamical trajectory of the excited-state probability demonstrate that Wei's TFSE has a significant simulation advantage in computational efficiency. Therefore, Wei's TFSE is more accurate and efficient for simulating the accelerated dynamics of non-Markovian quantum systems.
\end{abstract}





\maketitle

\section{Introduction}\label{sec:introduction}
The dynamical behavior of open quantum systems originates from the interaction between the system and its external environment~\cite{BreuerPetruccione2002, Lindblad1976}. Compared with a closed quantum system, the degrees of freedom in its environment introduce effects such as dissipation, decoherence, and information backflow, which often lead to complex system dynamics~\cite{Breueretal2016, Zurek2003, Weiss2012, BreuerLainePiilo2009}. The temporal behavior of an open quantum system weakly coupled to its environment is typically studied with the aid of the weak coupling approximation and the Markov approximation~\cite{BreuerPetruccione2002}. Specifically, the weak coupling approximation assumes that the interaction between the system and its environment is sufficiently weak, thereby simplifying the simulation of system dynamics~\cite{BreuerPetruccione2002, Davies1974}. The Markov approximation is based on the assumption that the correlation time of the environment is much shorter than the relaxation time of the system, so that the evolution of the system is mainly determined by its current state~\cite{Breueretal2016, Davies1974}. However, both approximations face several limitations and challenges when accurately describing the temporal behavior of a real physical system~\cite{Alickietal2004, Daffereretal2004, TerhalBurkard2005}. When the memory effects of the environment cannot be neglected, an open system exhibits the non-Markovian dynamical characteristics~\cite{AliferisGottesmanPreskill2006, Carusoetal2014}. Such dynamical behavior is of great significance in quantum information processing, including enhancing channel capacity~\cite{Meleetal2024}, accelerating quantum system evolution~\cite{Miaoetal2025}, realizing entanglement protocols~\cite{Gaikwadetal2024}, and describing quantum coherence in photosynthetic systems~\cite{Lorenzonietal2025}. Therefore, it is crucial to develop more accurate and efficient theories for simulating the non-Markovian dynamics.

Fractional calculus is regarded as a useful tool for describing the physical processes with memory effects and nonlocal characteristics~\cite{Iomin2009, Wuetal2010, Huangetal2011}, the Time-Fractional Schr\"odinger Equation (TFSE) offers a theoretical approach for simulating the non-Markovian quantum dynamical processes with memory effects~\cite{ZhaoLuo2019, Ertiketal2010, Laskin2017}. In the development of Time-Fractional Quantum Dynamics (TFQD), Naber proposed the TFSE based on the Caputo Fractional Derivative (Ca-FD), namely Naber's TFSE~\cite{Naber2004, Mainardi1996}. Related studies further obtained the exact solutions of this equation for systems such as free particles and particles in potential wells~\cite{ZuYu2022}. After its extension to the Time-Space-Fractional Schrödinger Equation, research on the free-particle problems in infinite square wells and $\delta$-potential wells were advanced~\cite{WangXu2007, DongXu2008}. These works indicate that Naber's TFSE can provide a new route for simulating the complex quantum dynamical processes. In addition, its application to the quantum comb model further demonstrates its ability to characterize the non-Markovian dynamical processes with memory effects~\cite{Iomin2009}.

Naber's TFSE provides a theoretical foundation for the study of the TFQD. However, existing results show that Naber's TFSE~\cite{Naber2004}, Naber's TFSE II~\cite{Naber2004}, and XGF's TFSE~\cite{XiangGuoFu2019} all violate probability conservation in quantum mechanics when describing the Time-Fractional Open Quantum System (TFOQS)~\cite{Weietal2024}. Meanwhile, these TFSEs are also restricted by the applicable interval of the fractional order $\beta$ when simulating the non-Markovian dynamics~\cite{Weietal2024}. In contrast, Wei's TFSE based on the Conformable Fractional Derivative (Co-FD) always preserves probability conservation and has shown the superior simulation performances in the Resonant Dissipative Jaynes-Cummings (RDJC) model and the double RDJC model. Furthermore, the ability of Wei's TFSE to maintain probability conservation can be extended to open quantum systems with the arbitrary number of qubits. In addition, the potential advantages of Wei's TFSE in simulating the non-Markovian quantum dynamical processes indicate that it will have important application values in open system models containing more qubits. From these two aspects, Naber's TFSE outperforms Naber's TFSE II and XGF's TFSE, but all three are inferior to Wei's TFSE~\cite{Weietal2024}.

The maximum evolution speed of a quantum system is a central issue across almost all areas of quantum physics~\cite{DeffnerCampbell2017}. It determines the operation speed of quantum gates, the operating efficiency of quantum computation, and the response time of quantum sensing~\cite{Taddeietal2013, Lloyd2000, HerbDegen2024}. The Quantum Speed Limit (QSL) time can quantify this speed and help reveal the essential characteristics in the evolution of a quantum system~\cite{Zhuetal2025}. Therefore, constructing a more accurate and efficient theoretical framework is of great significance for simulating the accelerated dynamics of the non-Markovian quantum systems. To this end, this paper employs Wei's TFSE to simulate the non-Markovian quantum accelerated evolution process in the RDJC model. The QSL time of a time-fractional single-qubit open system is solved and compared with the corresponding results obtained within the framework of Naber's TFSE~\cite{Weietal2023}. The main contributions are as follows:

(1) Within the framework of Wei's TFSE, several results consistent with those obtained under Naber's TFSE are found: (i) the enhancement mechanism of the system evolution speed induced by the non-Markovian memory effects of the environment is revealed; (ii) the optimized acceleration of the system evolution can be achieved by jointly regulating the fractional order, coupling strength, and photon number.

(2) Wei's TFSE can accurately capture the non-Markovian accelerated dynamical features of the system for all \mbox{$\beta\in(0,1]$}, whereas Naber's TFSE can do so only for large values of $\beta$.

(3) Wei's TFSE based on Co-FD has a significant simulation advantage in computational efficiency. Since Co-FD avoids the path integrals compared to Ca-FD, it reduces computational complexity.

The remainder of this paper is organized as follows. Section 2 introduces Naber's TFSE, Wei's TFSE, and the RDJC model. Section 3 derives the QSL time simulated by Wei's TFSE. Section 4 compares the simulation performances of Naber's TFSE and Wei's TFSE from four aspects. Finally, Section 5 summarizes the paper.

\section{Single-qubit Open System Simulated by TFSE}
\label{sec:Single-qubit Open System Simulated by TFSE}
This section introduces Naber's TFSE and Wei's TFSE, and then presents the RDJC model, providing the basis for calculating the QSL time.

\subsection{Naber's TFSE}
Naber's TFSE is based on the Ca-FD, which is defined as

\begin{equation}
\label{eq:1}
    {}_a^{Ca}D_t^\beta f(t) = \frac{1}{\Gamma(n-\beta)} \int_a^t \frac{f^{(n)}(\tau)}{(t-\tau)^{1-n+\beta}} d\tau,
\end{equation}

\noindent and is combined with the Analytic Continuation of Time (ACT) $t \to t/i\hbar_\beta$. Therefore,
the explicit form of Naber's TFSE is

\begin{equation}
\label{eq:2}
    (i\hbar_\beta)^\beta \frac{\partial^\beta \left|\psi(x,t)\right\rangle}{\partial t^\beta} = H_\beta \left|\psi(x,t)\right\rangle,
\end{equation}

\noindent where $\beta\in(0,1]$, and $H_\beta$ is the pseudo-Hamiltonian.

\subsection{Wei's TFSE}
Wei's TFSE adopts the Co-FD, which is given by

\begin{equation}
\label{eq:3}
    {}_a^{\mathrm{Co}}D_t^\beta f(t) = \lim_{\varepsilon\to 0}\frac{f\left(t+\varepsilon t^{1-\beta}\right)-f(t)}{\varepsilon},
\end{equation}

\noindent combined with the ACT $t \to t/(i\hbar_\beta\beta)^{1/\beta}$. Thus, the explicit form of Wei's TFSE is

\begin{equation}
    i\hbar_\beta {}_0^{\mathrm{Co}}D_t^\beta \left|\psi(x,t)\right\rangle = H_\beta \left|\psi(x,t)\right\rangle.
\end{equation}

\subsection{RDJC Model}
The RDJC model is used to characterize the resonant coupling dynamics of a two-level system interacting with its own dissipative environment at zero temperature. Under the rotating-wave approximation, assuming the Planck constant $\hbar=1$, the total Hamiltonian $H_{\mathit{Total}}$ describing the interaction between the single-qubit system and its dissipative environment is written as

\begin{equation}
    H_{\mathit{Total}} = H_S + H_E + H_I,
\end{equation}

\noindent where

\begin{equation}
    H_S = \omega_0^S \sigma_+^S \sigma_-^S,
\end{equation}

\begin{equation}
    H_E = \sum_{j\in\mathbb{N}^+} \omega_j^E (b_j^E)^\dagger b_j^E.
\end{equation}

\noindent Here, $H_S$ is the free Hamiltonian of the qubit system, and $H_E$ is the free Hamiltonian of the environment. $\omega_0^S$ denotes the transition frequency of the two-level system between the excited state and the ground state. $\sigma_\pm^S$ are the atomic raising and lowering Pauli operators. $\omega_j^E$ and $(b_j^E)^\dagger b_j^E$ correspond to the angular frequency and the creation (annihilation) operators of the $j$-th mode of the cavity field, respectively.

$H_I$ is the interaction Hamiltonian between the two-level system and its environment, expressed as

\begin{equation}
    H_I = \sum_{j\in\mathbb{N}^+} \lambda_j^{SE} \big(\sigma_-^S (b_j^E)^\dagger + \sigma_+^S b_j^E\big),
\end{equation}

\noindent where $\lambda_j^{SE}\in[0,1]$ denotes the coupling strength.

To treat the interaction process between the system and the environment more effectively, a time-dependent unitary transformation is applied to $H_{\mathit{Total}}$, transforming the system into the interaction picture. Under the resonance condition, $H_I$ takes the form

\begin{equation}
    H_I(t) = \sum_{j\in\mathbb{N}^+} \lambda_j \big(\sigma_-^S (b_j^E)^\dagger e^{-i(\omega_0-\omega_j)t} + \sigma_+^S b_j^E e^{i(\omega_0-\omega_j)t}\big).
\end{equation}

\section{QSL time Simulated by Wei's TFSE}
\label{sec:QSL time Simulated by Wei's TFSE}
The QSL time defines the minimum time required for a quantum system to evolve from an initial state to a target state. This time is crucial for quantifying the upper bound of the evolution speed in quantum processes. In this section, the Margolus-Levitin-type bound~\cite{DeffnerLutz2013} based on a pure initial state is adopted to calculate the QSL time of the single-qubit open system under the framework of Wei's TFSE. The formula is given by

\begin{equation}
\label{eq:10}
    \tau_{QSL} = \frac{\sin^2\big[B(\rho_S(0),\rho_S(\tau))\big]}{\Lambda_\tau^{op}}
\max\left\{\frac{1}{\Lambda_\tau^{tr}},\frac{1}{\Lambda_\tau^{hs}},\frac{1}{\Lambda_\tau^{op}}\right\},
\end{equation}

\noindent where $B(\rho_S(0),\rho_S(\tau))=\arccos\sqrt{\langle\psi(0)|\rho_S(\tau)|\psi(0)\rangle}$ denotes the Bures angle between the initial state and the target state of the open quantum system, which is used to characterize the distinguishability and information retention degree of the two quantum states. $\Lambda_\tau^p=\frac{1}{\tau}\int_0^\tau \|\dot{\rho}_S(t)\|_p dt$ denotes the average value of $\|\dot{\rho}_S(t)\|_p$ over the driving time $\tau\in[0,1]$. $\|A\|_p=\big(s_1^p+s_2^p+\cdots+s_n^p\big)^{1/p}$ is the Schatten $p$-norm, where $p=tr$, $p=hs$, and $p=op$ are the trace norm, Hilbert-Schmidt norm, and operator norm, respectively. The norms of $\|A\|_p$ are defined as

\begin{equation}
\label{eq:11}
\begin{aligned}
\|A\|_{tr} &= \sum_{m} s_m, \\
\|A\|_{hs} &= \sqrt{\sum_{m} s_m^2}, \\
\|A\|_{op} &= \max_{m} \{s_m\},
\end{aligned}
\end{equation}

\noindent where $s_1,s_2,\dots,s_n$ are the singular values of $A$.

When the time-fractional single-qubit open system evolves from the pure initial state $\rho_S^1(0)=|e\rangle\langle e|$ to the target state $\rho_{S_4}^1(\tau)$, the exact solution of the single-qubit open system under the framework of Wei's TFSE can be obtained from Ref.~\cite{Weietal2024},

\begin{equation}
\label{eq:12}
\begin{aligned}
\rho_{S}^{1(11)} &= \left(e^2_{\beta_2^+} - e^2_{\beta_2^-}\right)\left(e^{2*}_{\beta_2^+} - e^{2*}_{\beta_2^-}\right)a'^2 b'^2, \\
\rho_{S}^{1(22)} &= \left(e^2_{\beta_2^+} + e^2_{\beta_2^-}\right)\left(e^{2*}_{\beta_2^+} + e^{2*}_{\beta_2^-}\right)b'^4.
\end{aligned}
\end{equation}

First, one can obtain

\begin{equation}
\begin{aligned}
&\sin^2\left[B\left(\rho_S^1(0),\rho_S^1(\tau)\right)\right] \\
&= \left|\operatorname{tr}\left(\rho_S^1(0)\rho_S^1(\tau)\right)-1\right| \\
&= \left|
\frac{
a^2
\left(e^2_{\beta_2^-}-e^2_{\beta_2^+}\right)
\left(e^{2*}_{\beta_2^-}-e^{2*}_{\beta_2^+}\right)
}{
\begin{aligned}
&(b^2-a^2)
\left(
e^2_{\beta_2^-}e^{2*}_{\beta_2^+}
+e^2_{\beta_2^+}e^{2*}_{\beta_2^-}
\right) \\
&-(b^2+a^2)
\left(
e^2_{\beta_2^-}e^{2*}_{\beta_2^-}
+e^2_{\beta_2^+}e^{2*}_{\beta_2^+}
\right)
\end{aligned}
}
\right|.
\end{aligned}
\end{equation}

\noindent Then, all singular values $s_1,s_2,\dots,s_p$ ($p\in\mathbb{N}^+$) of $\dot{\rho}_S(\tau)$ are calculated. Based on the operator norm expression of the non-unitary generator in Eq.~\eqref{eq:11}, the largest singular value among all singular values is obtained as $s_{\mathrm{max}}=\|\dot{\rho}_{S4}^1(\tau)\|_{op}$. Therefore, for the time-fractional single-qubit open system, the QSL time in Eq.~\eqref{eq:10} can finally be expressed as

\begin{equation}
\begin{aligned}
\frac{\tau_{QSL}}{\tau}
&= \frac{1}{\int_{0}^{\tau} s_{\max}\,dt}
\left|
\frac{
a^2
\left(e^2_{\beta_2^-}-e^2_{\beta_2^+}\right)
\left(e^{2*}_{\beta_2^-}-e^{2*}_{\beta_2^+}\right)
}{
\begin{aligned}
&(b^2-a^2)
\left(
e^2_{\beta_2^-}e^{2*}_{\beta_2^+}
+e^2_{\beta_2^+}e^{2*}_{\beta_2^-}
\right) \\
&-(b^2+a^2)
\left(
e^2_{\beta_2^-}e^{2*}_{\beta_2^-}
+e^2_{\beta_2^+}e^{2*}_{\beta_2^+}
\right)
\end{aligned}
}
\right|.
\end{aligned}
\end{equation}

The above derivation clearly shows that the QSL time of the time-fractional single-qubit open system is affected not only by the state parameters $a$, $b$ and the evolution time $\tau$, but is also constrained by three key parameters: the fractional order $\beta$, the coupling strength $\lambda$, and the photon number $n$. To simplify the analysis, the state parameters are set as $a=b=\sqrt{2}/2$ in this section. In evaluating the maximum evolution speed of a quantum system, the QSL time ratio $\tau_{QSL}/\tau$ plays an important role. When $\tau_{QSL}/\tau = 1$, the quantum system has completed the target evolution within its speed limit time, indicating that the evolution speed reaches its limit and that the system can no longer undergo further accelerated evolution. When $\tau_{QSL}/\tau < 1$, the evolution speed of the quantum system has not yet reached its limit, indicating the potential for further accelerated evolution. Specifically, a smaller value of $\tau_{QSL}/\tau$ corresponds to a stronger potential acceleration capability of the quantum system. The following numerical analyses further investigate how these three key parameters affect the QSL time of the time-fractional single-qubit open system, so as to reveal the underlying evolution mechanism.

\section{Performance Comparisons of Two TFSEs in Simulating Non-Markovian Quantum Accelerated Dynamics}
\label{sec:Performance Comparisons of Two TFSEs in Simulating Non-Markovian Quantum Accelerated Dynamics}
\begin{figure*}[!t]
    \centering
    \includegraphics[width=0.9\textwidth,
        trim=0 100 0 95,
        clip
        ]{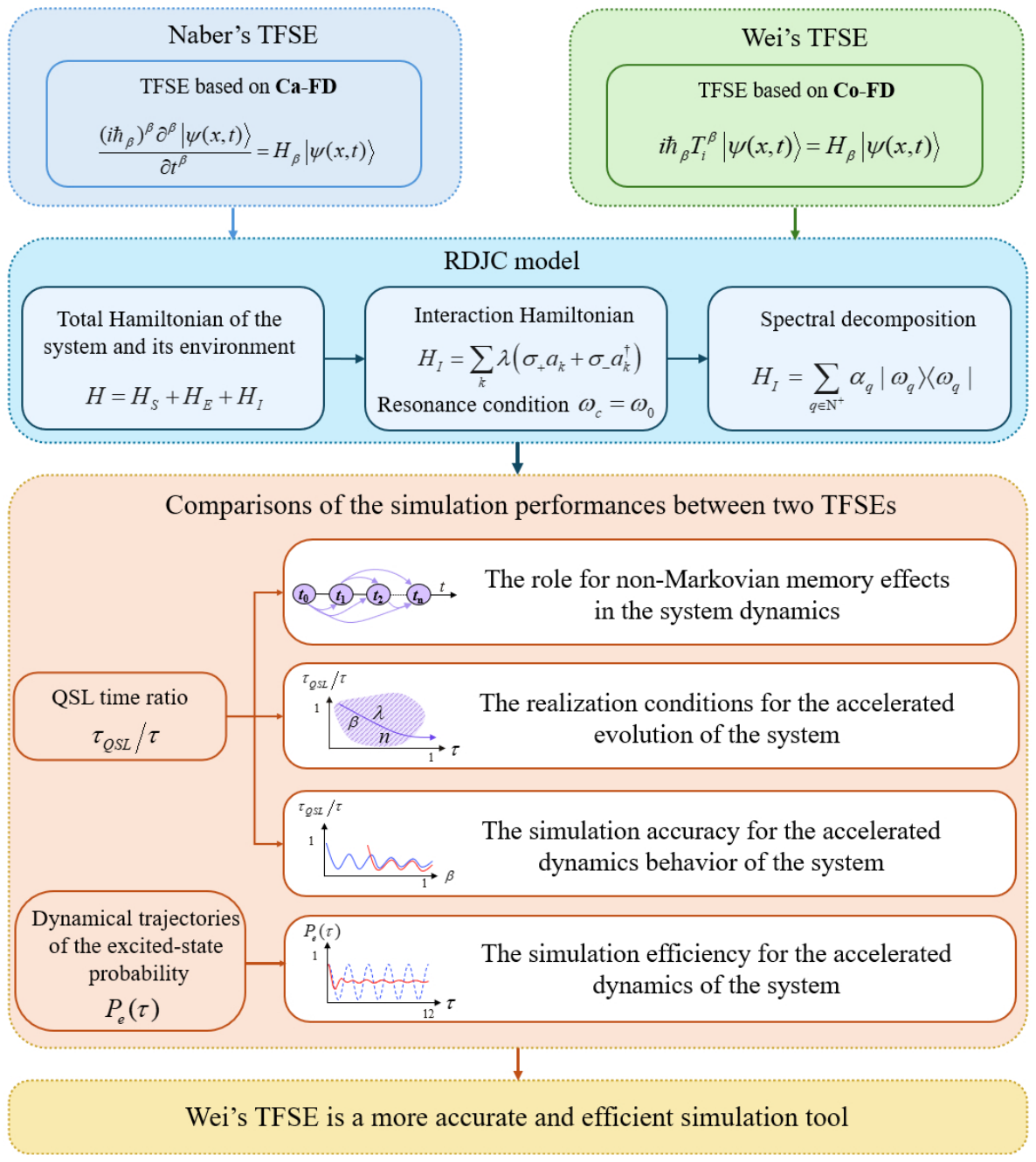}
    \caption{Schematic framework of this work. The simulation performances of Naber's TFSE and Wei's TFSE in the RDJC model are compared by calculating the QSL time ratio $\tau_{QSL}/\tau$ and tracking the dynamical trajectory of the excited-state probability $P_e(\tau)$.}\label{fig:framework}
\end{figure*}

This section investigates the performances of Wei's TFSE in simulating non-Markovian quantum accelerated dynamics and compares it in detail with the results simulated by Naber's TFSE. The comparison is conducted from the four aspects outlined in Fig.~\ref{fig:framework}.

\subsection{Role for the Non-Markovian Memory Effects in System Dynamics}
Existing studies have shown that, under the framework of Naber's TFSE, the non-Markovian memory effects of the environment can increase the evolution speed of the system~\cite{Weietal2023}. Therefore, the influence of the non-Markovian memory effects on system dynamics under the framework of Wei's TFSE is considered.

\begin{figure}[pos=h]
    \centering

    \begin{minipage}[b]{0.49\linewidth}
        \centering
        \includegraphics[width=\linewidth]{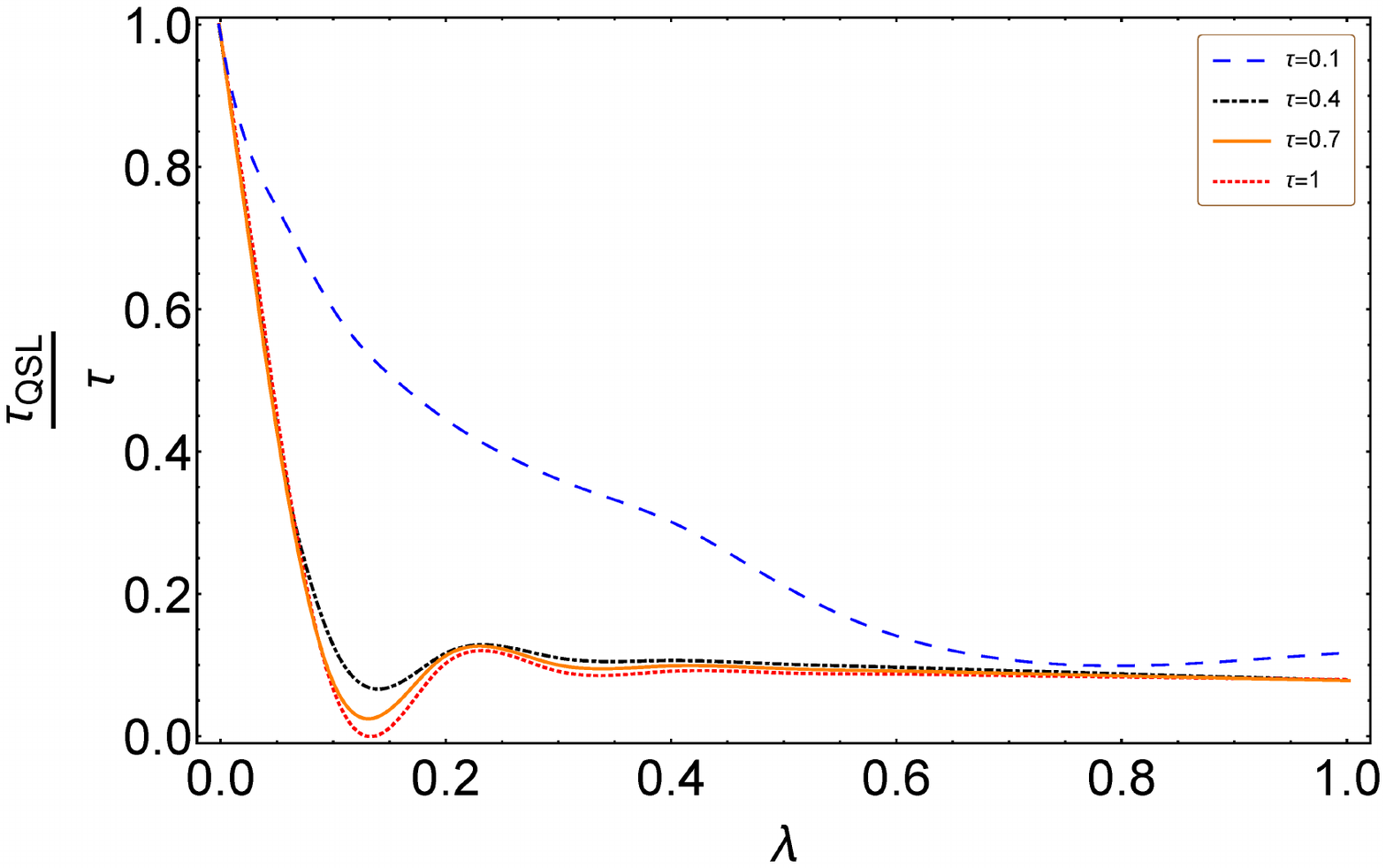}
        \makebox[\linewidth][c]{\small (a) $\beta=0.25$}
    \end{minipage}
    \hfill
    \begin{minipage}[b]{0.49\linewidth}
        \centering
        \includegraphics[width=\linewidth]{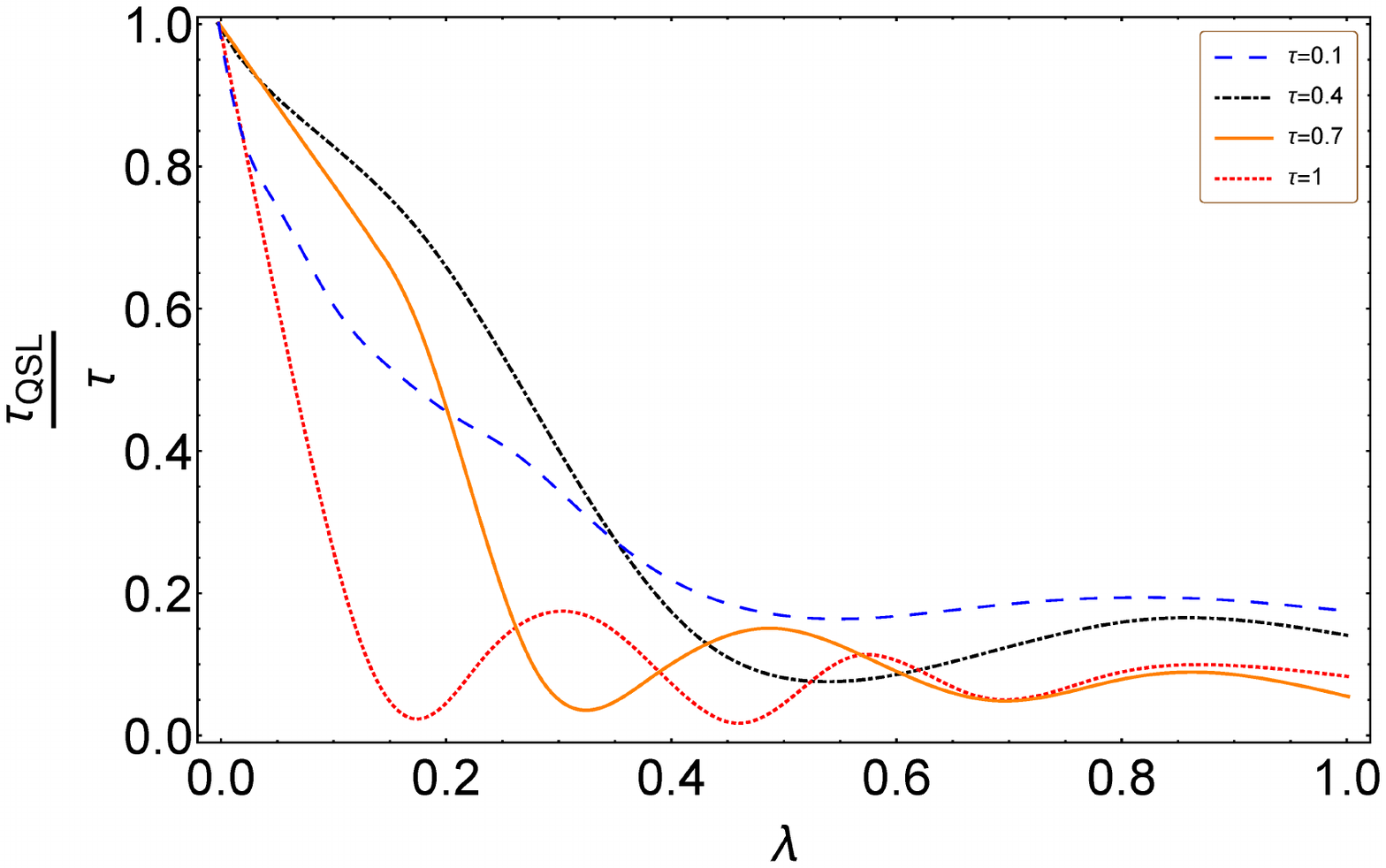}
        \makebox[\linewidth][c]{\small (b) $\beta=0.5$}
    \end{minipage}

    \begin{minipage}[b]{0.49\linewidth}
        \centering
        \includegraphics[width=\linewidth]{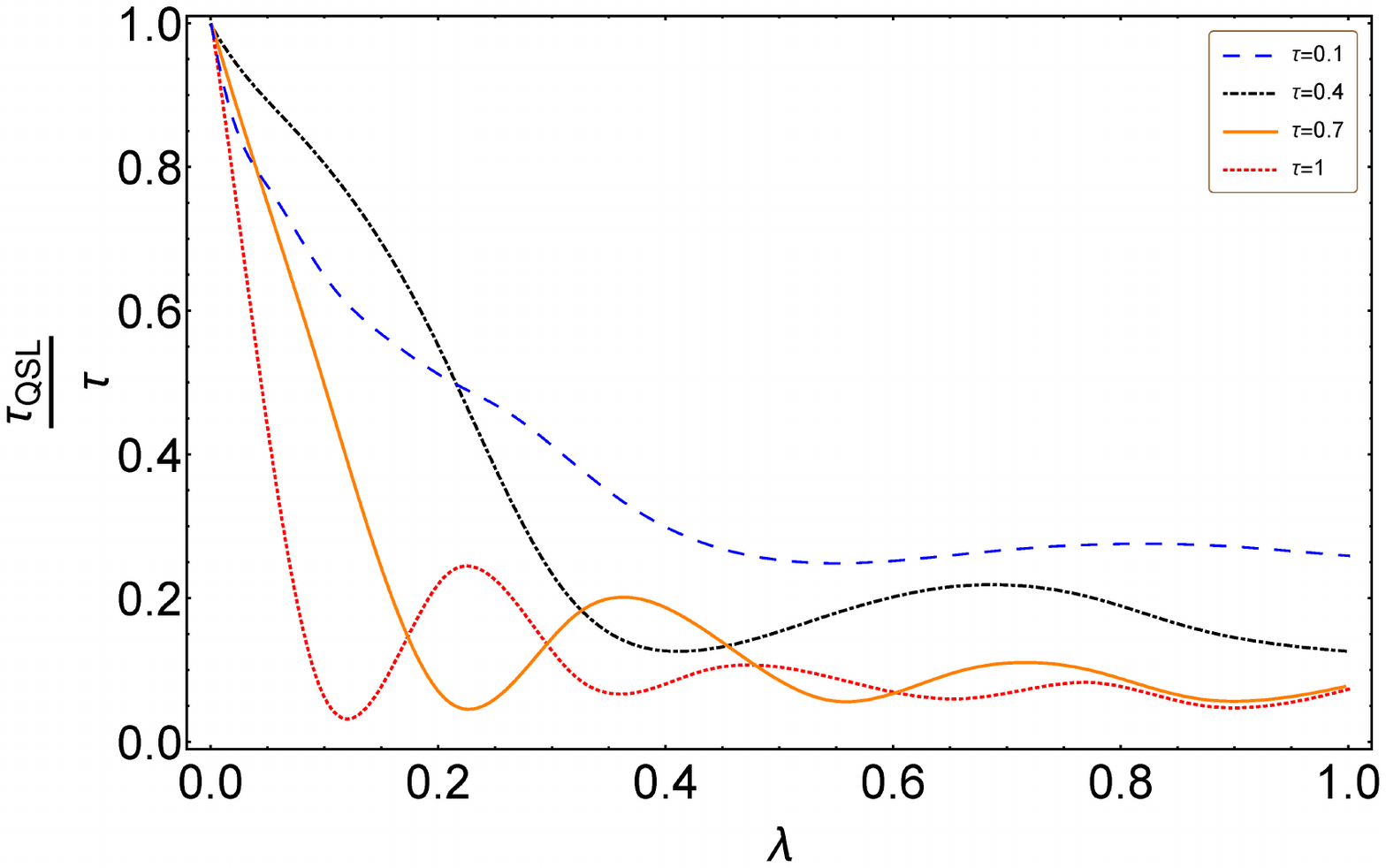}
        \makebox[\linewidth][c]{\small (c) $\beta=0.75$}
    \end{minipage}
    \hfill
    \begin{minipage}[b]{0.49\linewidth}
        \centering
        \includegraphics[width=\linewidth]{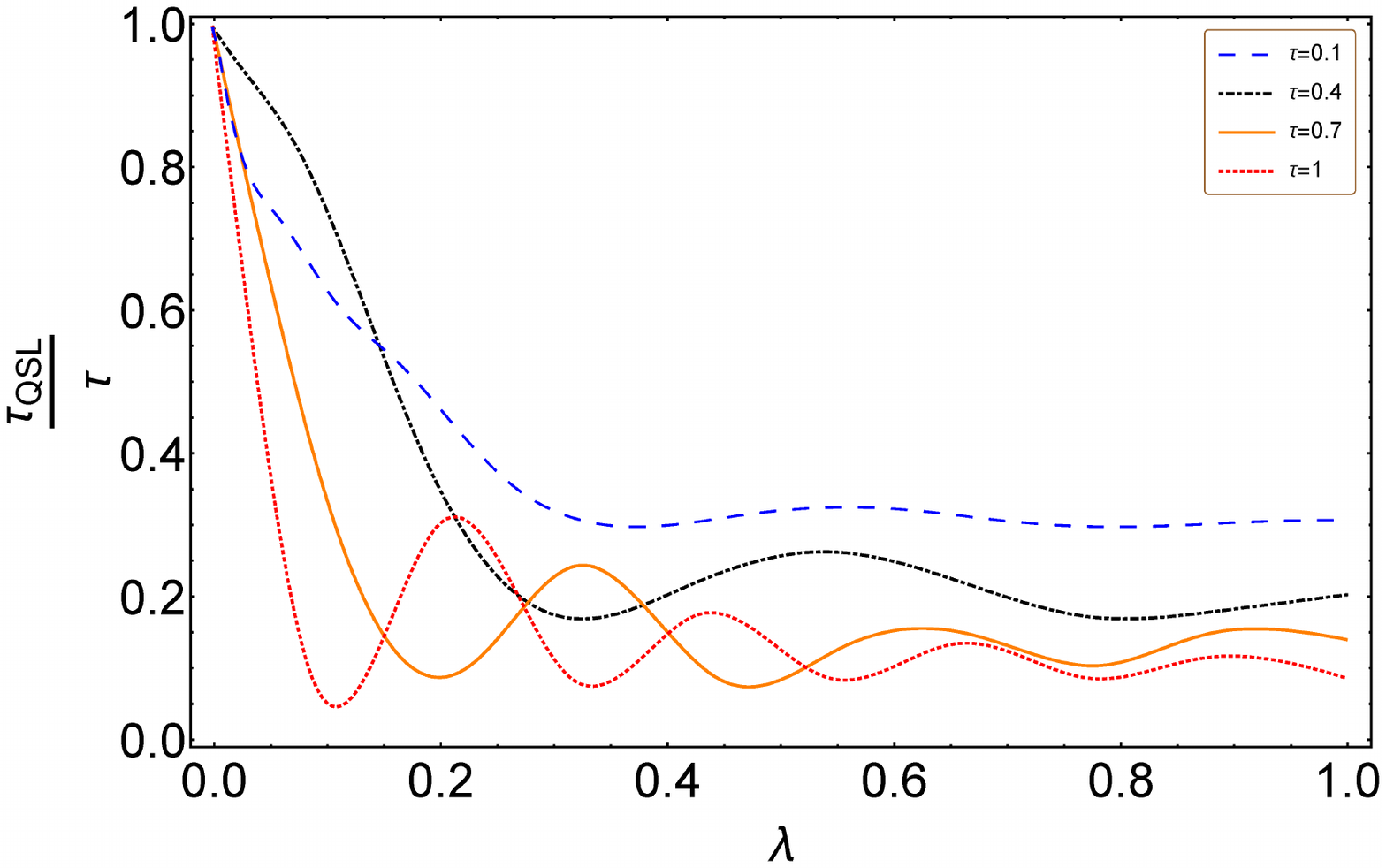}
        \makebox[\linewidth][c]{\small (d) $\beta=1$}
    \end{minipage}

    \caption{Under the framework of Wei's TFSE, the QSL time ratio $\tau_{QSL}/\tau$ of the single-qubit open system as a function of the coupling strength $\lambda$ for different fractional orders $\beta$ and evolution times $\tau$, where (a) $\beta=0.25$, (b) $\beta=0.5$, (c) $\beta=0.75$, and (d) $\beta=1$. Other parameters are set as $\tau=0.1,0.4,0.7,1$ and $n=40$.}
    \label{fig:4.1}
\end{figure}

Fig.~\ref{fig:4.1} shows the variation of the $\tau_{QSL}/\tau$ with $\lambda$ for a single-qubit open system under the framework of Wei's TFSE, where different fractional orders $\beta=0.25,0.5,0.75,1$ and evolution times $\tau=0.1,0.4,0.7,1$ are selected, and the photon number is set as $n=40$. As directly observed from Fig.~\ref{fig:4.1}(a), even when both $\beta$ and $\tau$ take relatively small values, accelerated evolution of the system remains evident and common under the framework of Wei's TFSE. However, under the framework of Naber's TFSE, this acceleration phenomenon is not evident when both $\beta$ and $\tau$ remain small. In the description based on Wei's TFSE, when $\tau$ is fixed, the non-Markovian oscillatory behavior of $\tau_{QSL}/\tau$ becomes more pronounced as $\beta$ gradually increases from small values, with both the oscillation amplitude and frequency showing clear enhancement. This indicates that, as $\beta$ increases, the evolution speed of the TFOQS becomes faster, while the potential for further accelerated evolution of the system decreases. In addition, Fig.~\ref{fig:4.1}(a-b) clearly shows that, on the short time scales, $\tau_{QSL}/\tau$ already exhibits non-Markovian oscillatory behavior at small $\beta$ values. Compared with the dynamical behavior on short time scales, Fig.~\ref{fig:4.1}(c-d) shows that $\tau_{QSL}/\tau$ exhibits more pronounced non-Markovian oscillations in the long-time dynamics. These oscillations indicate that the non-Markovian memory effects of the environment can accelerate the evolution of the TFOQS, thereby leading to a smaller QSL time. The same conclusion is also obtained under the framework of Naber's TFSE.

\subsection{Realization Conditions for the Accelerated Evolution of the System}
\label{sec:Realization Conditions for the Accelerated Evolution of the System}
Under the framework of Naber's TFSE, the conditions for realizing accelerated evolution of the system depend on the coordinated regulation among $\beta$, $\lambda$, and $n$. Specifically, when both $\lambda$ and $n$ are small while $\beta$ is large, the system has a lower acceleration potential during its evolution in the memory environment. In this case, the evolution speed of the system is faster~\cite{Weietal2023}. This subsection further analyzes the influence of these three parameters on the accelerated evolution of the system under the framework of Wei's TFSE.

\begin{figure}[pos=h]
\centering
\includegraphics[width=0.9\linewidth]{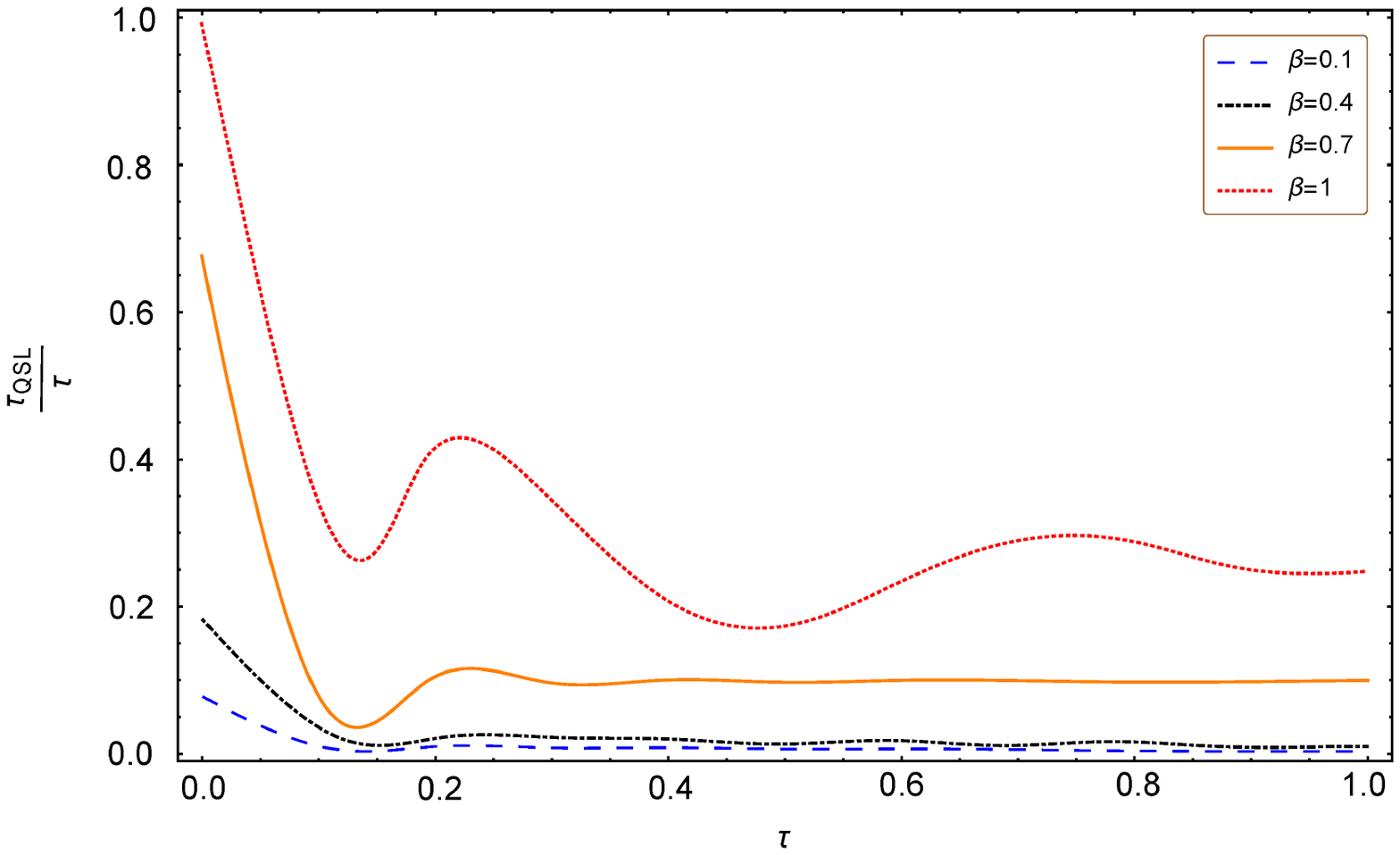}
\caption{Under the framework of Wei's TFSE, the QSL time ratio $\tau_{QSL}/\tau$ of the single-qubit open system as a function of the evolution time $\tau$ for different fractional orders $\beta=0.1,0.4,0.7,1$, where the coupling strength is set as $\lambda=0.5$ and the photon number as $n=40$.}
\label{fig:4.21}
\end{figure}

Fig.~\ref{fig:4.21} plots the relationship between $\tau_{QSL}/\tau$ and $\beta$ for the single-qubit open system under the framework of Wei's TFSE, where $\lambda=0.5$, $n=40$, and different fractional orders $\beta=0.1,0.4,0.7,1$ are selected. As shown in Fig.~\ref{fig:4.21}, $\tau_{QSL}/\tau$ clearly depends on $\beta$. Notably, even when $\beta$ is particularly small, $\tau_{QSL}/\tau$ still exhibits clear non-Markovian oscillatory behavior. This finding is in sharp contrast to the results obtained under the framework of Naber's TFSE, as shown in Table~\ref{tab:1}. Further observation shows that as the value of $\beta$ gradually increases, both the oscillation frequency and amplitude of $\tau_{QSL}/\tau$ become stronger. When $\beta$ is sufficiently large, such non-Markovian oscillatory behavior persists as $\tau$ increases. This indicates that the non-Markovian dissipative dynamics of the TFOQS can still be effectively described and interpreted on long time scales. It should be emphasized that when a larger $\beta$ value is selected, the QSL time of the system shortens and the evolution speed of the system becomes faster. In other words, a smaller $\beta$ value induces a stronger potential for the accelerated evolution of the system. This conclusion is consistent with the results obtained under the framework of Naber's TFSE~\cite{Weietal2023}.

\begin{table}[width=\linewidth,cols=3,pos=h]
\centering
\caption{Comparisons of the influence of the fractional order $\beta$, coupling strength $\lambda$, and photon number $n$ on the non-Markovian evolution mode of the system under Naber's TFSE and Wei's TFSE frameworks}
\label{tab:1}
\small

\begin{tabularx}{\tblwidth}{@{}
  >{\centering\arraybackslash}c
  >{\centering\arraybackslash}X
  >{\centering\arraybackslash}X
@{}}
\toprule
Parameters & Under Naber's TFSE framework, the system evolves in a non-Markovian manner & Under Wei's TFSE framework, the system evolves in a non-Markovian manner
\\
\midrule
Fractional order $\beta$ & $\beta \in (\tilde{\beta},1]$ & $\beta \in (0,1]$ \\
Coupling strength $\lambda$ & $\lambda \in [0,1]$ & $\lambda \in [0,1]$ \\
Photon number $n$ & $n \in \mathbb{N}$ & $n \in \mathbb{N}$ \\
\bottomrule
\end{tabularx}

\smallskip
\begin{minipage}{\tblwidth}
\small
\noindent Note: $\tilde{\beta}\in(0,1)$ denotes the critical point under the framework of Naber's TFSE at which, under the influence of $\beta$, the system evolution changes from Markovian to non-Markovian.
\end{minipage}
\end{table}

\begin{figure}[pos=h]
\centering
\includegraphics[width=0.9\linewidth]{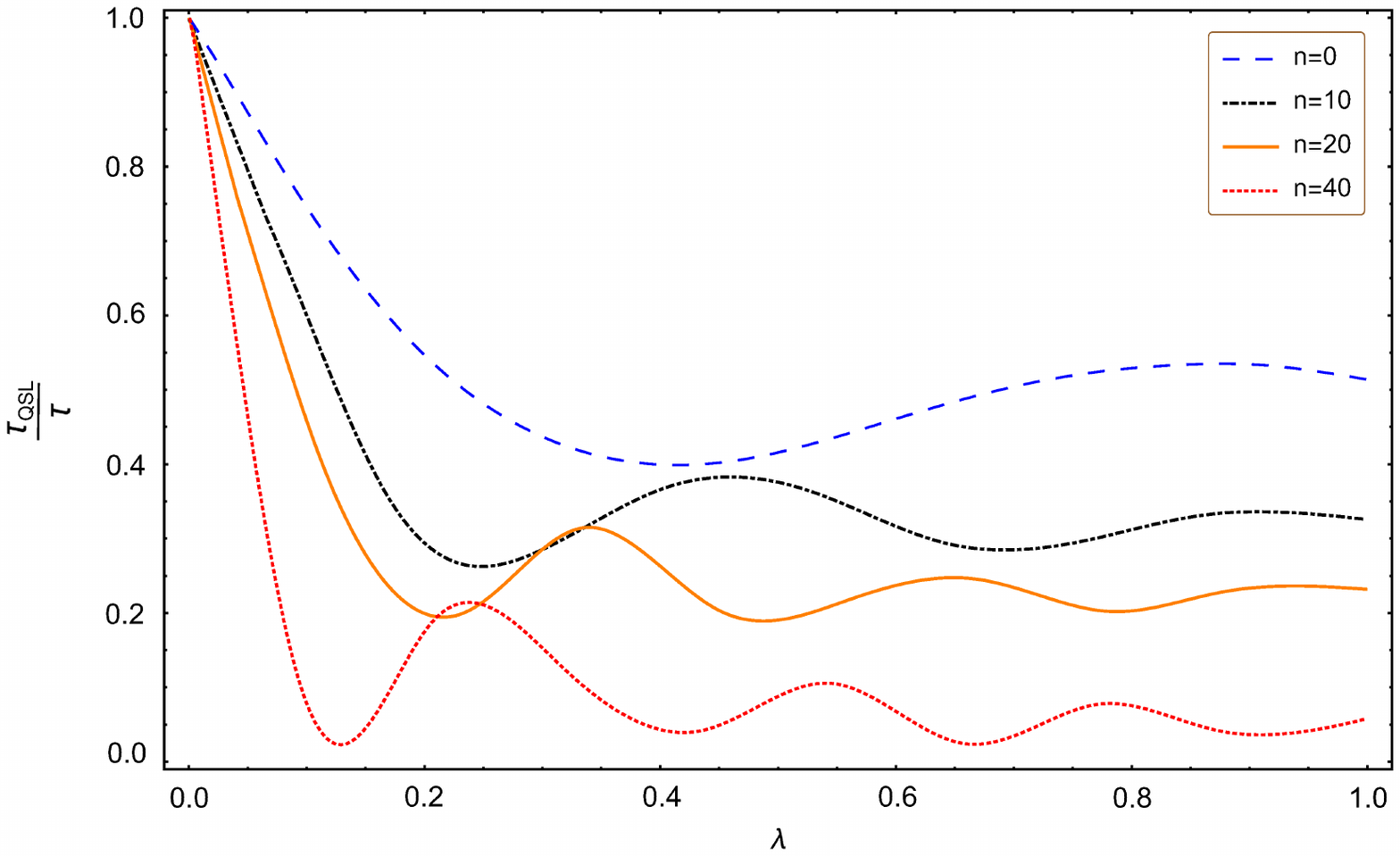}
\caption{Under the framework of Wei's TFSE, the QSL time ratio $\tau_{QSL}/\tau$ of the single-qubit open system as a function of the coupling strength $\lambda$ for different photon numbers $n=0,10,20,40$, where the fractional order is set as $\beta=0.8$ and the evolution time as $\tau=1$.}
\label{fig:4.22}
\end{figure}

Fig.~\ref{fig:4.22} shows the variation of $\tau_{QSL}/\tau$ with the coupling strength $\lambda$ for the single-qubit open system under the framework of Wei's TFSE, where $\tau=1$ is fixed and different photon numbers $n=0,10,20,40$ are selected. It is clear from Fig.~\ref{fig:4.22} that, even when $n=0$, the non-Markovian oscillatory behavior of $\tau_{QSL}/\tau$ remains evident. Moreover, as $n$ gradually increases, the value of $\tau_{QSL}/\tau$ shows a decreasing trend, accompanied by a higher oscillation frequency, while the change in amplitude is relatively moderate. This indicates that the non-Markovian memory effects of the environment can increase the evolution speed of the TFOQS, thereby shortening the QSL time required for it to reach the target state. Furthermore, as $\lambda$ increases, the value of $\tau_{QSL}/\tau$ also shows a decreasing trend. Based on the above observations and analysis, it can be concluded that the evolution of the TFOQS will obtain greater acceleration potential by reducing $\beta$ and correspondingly increasing $\lambda$ and $n$. In other words, when both $\lambda$ and $n$ are small while $\beta$ is large, the evolution speed of the TFOQS will be faster. This finding is consistent with the results obtained under the framework of Naber's TFSE and further clarifies an effective strategy for realizing accelerated the evolution of the system. Table~\ref{tab:2} lists in detail the specific influence of these parameters on the acceleration potential of the system, providing clear guidance for optimizing the evolution performances of the system.

\begin{table}[width=\linewidth,cols=3,pos=h]
\centering
\caption{Comparisons of the acceleration potential of the system under Naber's TFSE and Wei's TFSE frameworks}
\label{tab:2}
\small
\renewcommand{\arraystretch}{1.35}

\begin{tabularx}{\tblwidth}{@{}
  >{\centering\arraybackslash}c
  >{\centering\arraybackslash}X
  >{\centering\arraybackslash}X
@{}}
\toprule
Parameters & Naber's TFSE & Wei's TFSE \\
\midrule
Fractional order $\beta$ & $\beta \downarrow \,,\quad C_{1} \uparrow$ & $\beta \downarrow \,,\quad C_{2} \uparrow$ \\
Coupling strength $\lambda$ & $\lambda \uparrow \,,\quad C_{1} \uparrow$ & $\lambda \uparrow \,,\quad C_{2} \uparrow$ \\
Photon number $n$ & $n \uparrow \,,\quad C_{1} \uparrow$ & $n \uparrow \,,\quad C_{2} \uparrow$ \\
\bottomrule
\end{tabularx}

\begin{minipage}{\tblwidth}
\small
\noindent Note: $C$ denotes the potential for accelerated evolution, $\uparrow$ denotes an increase, and $\downarrow$ denotes a decrease.
\end{minipage}
\end{table}

\subsection{Simulation Accuracy for the Accelerated Dynamical Behavior of the System}
\label{sec:Simulation Accuracy for the Accelerated Dynamical Behavior of the System}
\begin{figure}[pos=h]
    \centering

    \begin{minipage}[b]{0.49\linewidth}
        \centering
        \includegraphics[width=\linewidth]{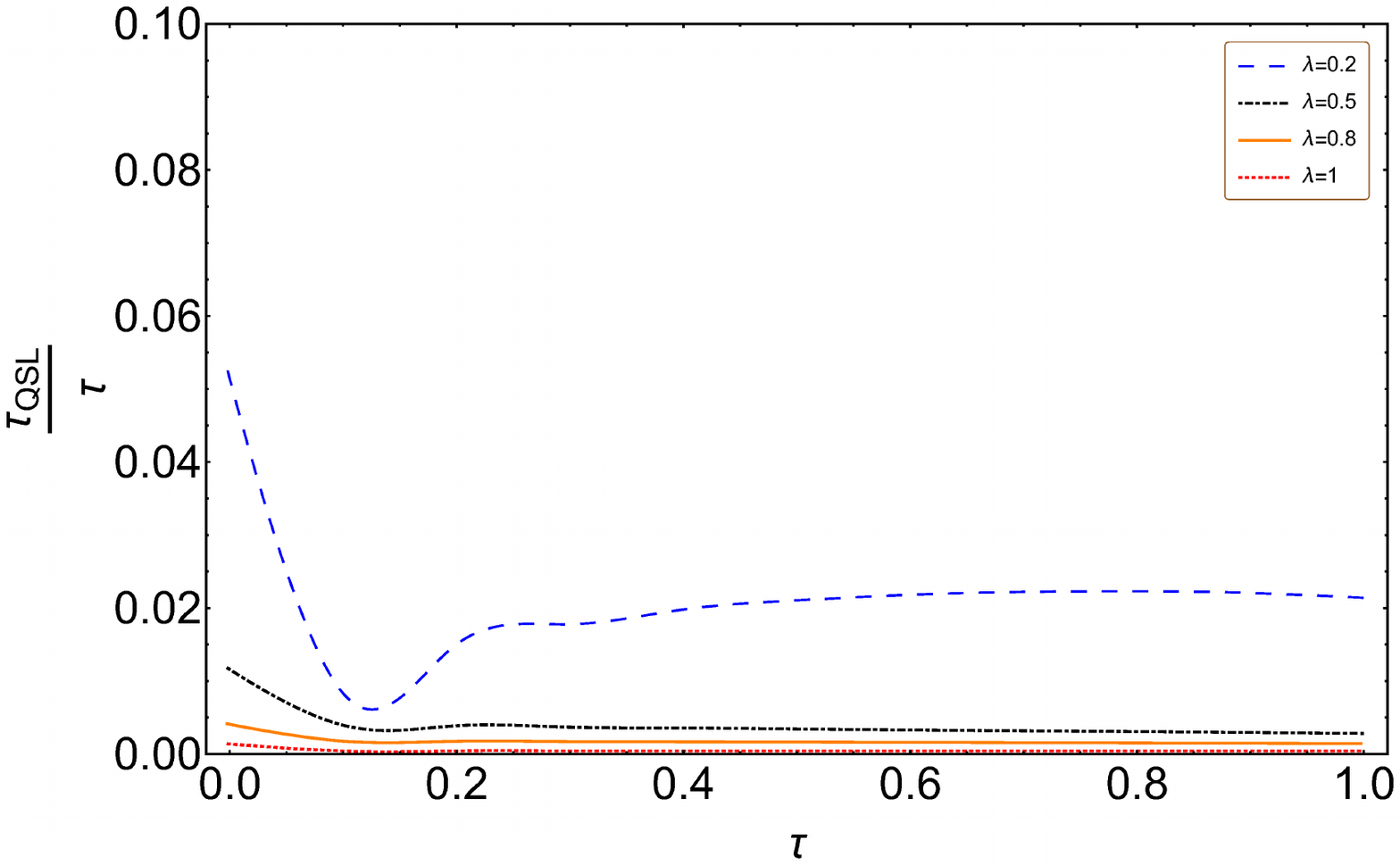}
        \makebox[\linewidth][c]{\small (a1) $\beta=0.05$}
    \end{minipage}
    \hfill
    \begin{minipage}[b]{0.49\linewidth}
        \centering
        \includegraphics[width=\linewidth]{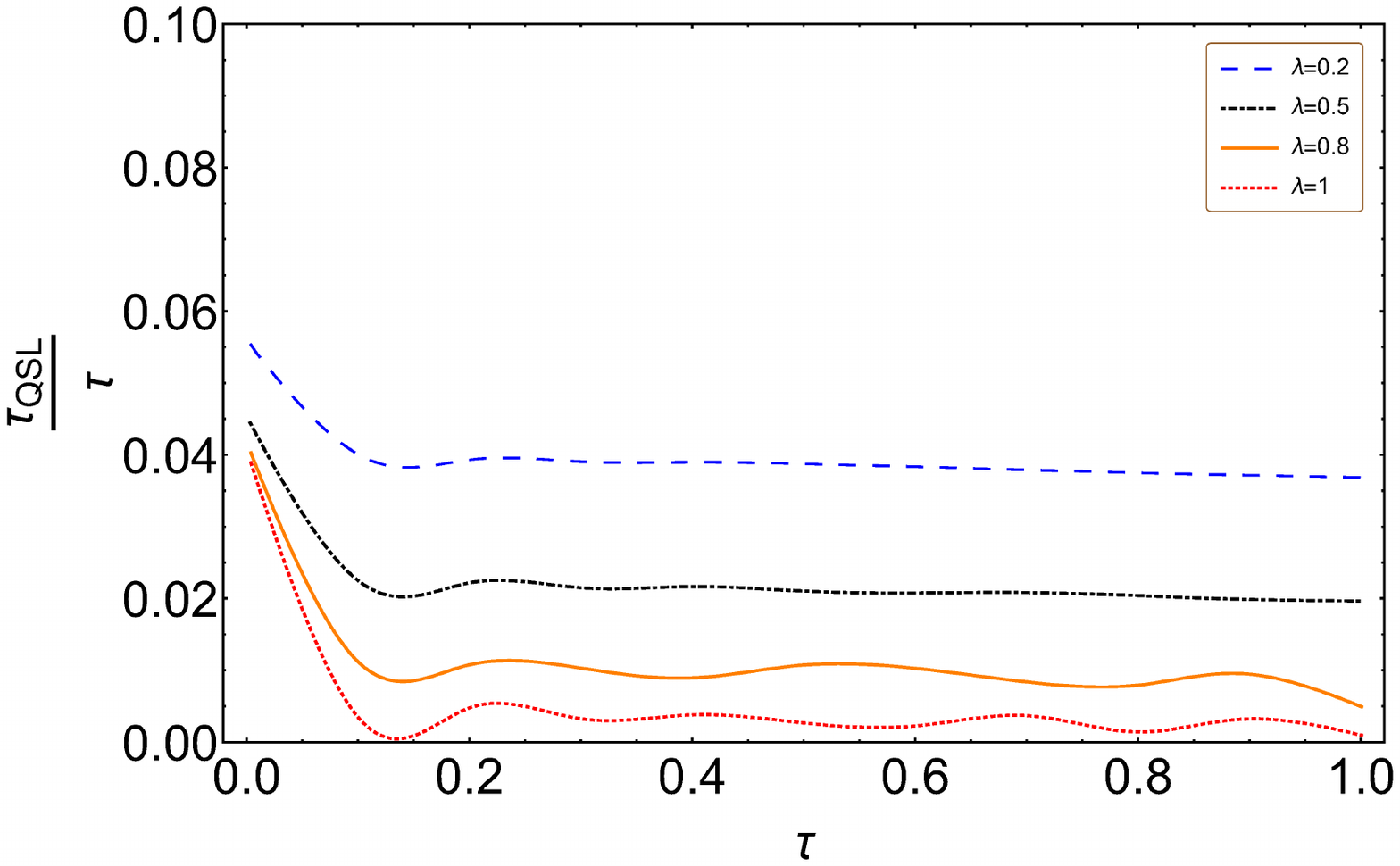}
        \makebox[\linewidth][c]{\small (a2) $\beta=0.05$}
    \end{minipage}

    \begin{minipage}[b]{0.49\linewidth}
        \centering
        \includegraphics[width=\linewidth]{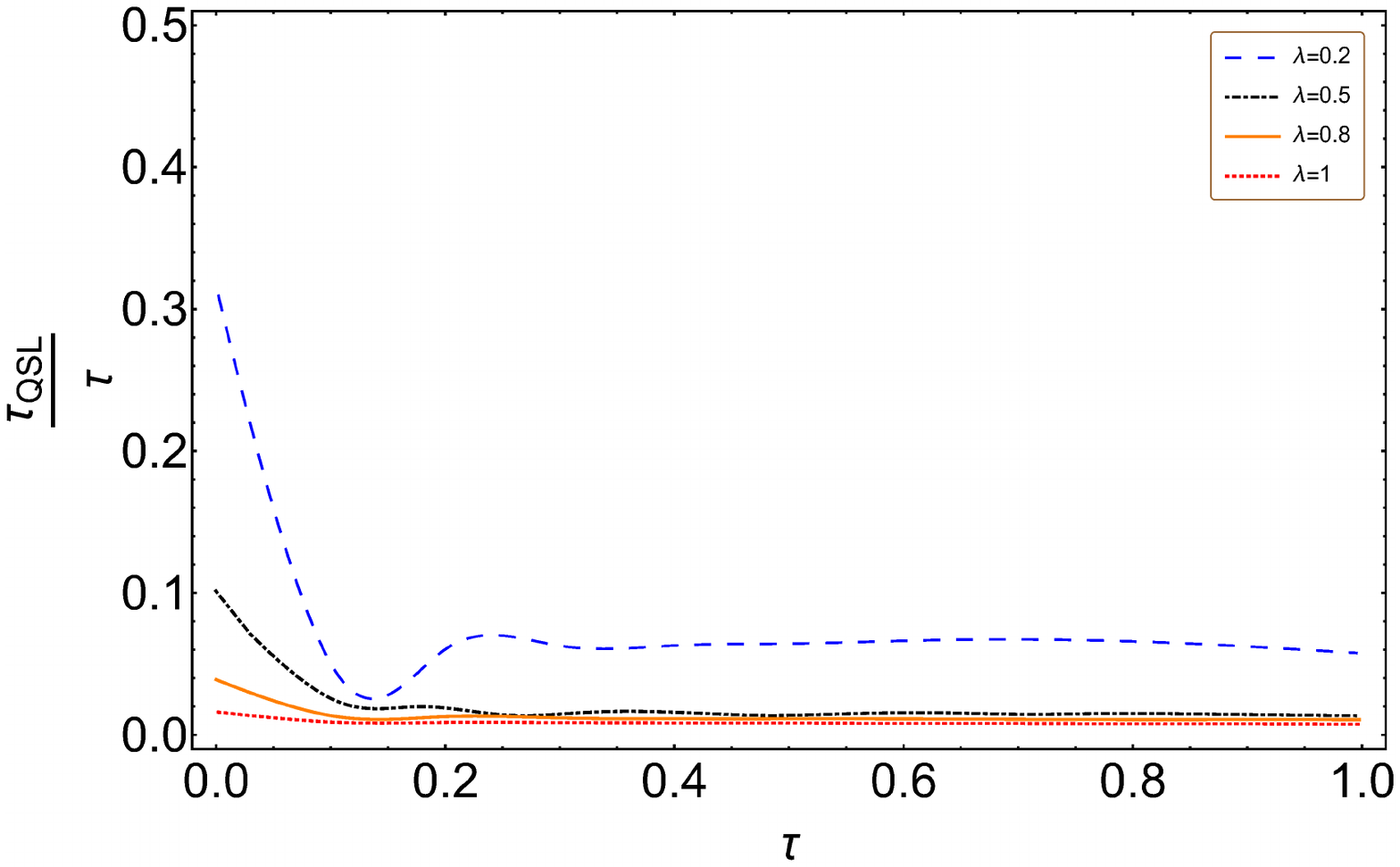}
        \makebox[\linewidth][c]{\small (b1) $\beta=0.35$}
    \end{minipage}
    \hfill
    \begin{minipage}[b]{0.49\linewidth}
        \centering
        \includegraphics[width=\linewidth]{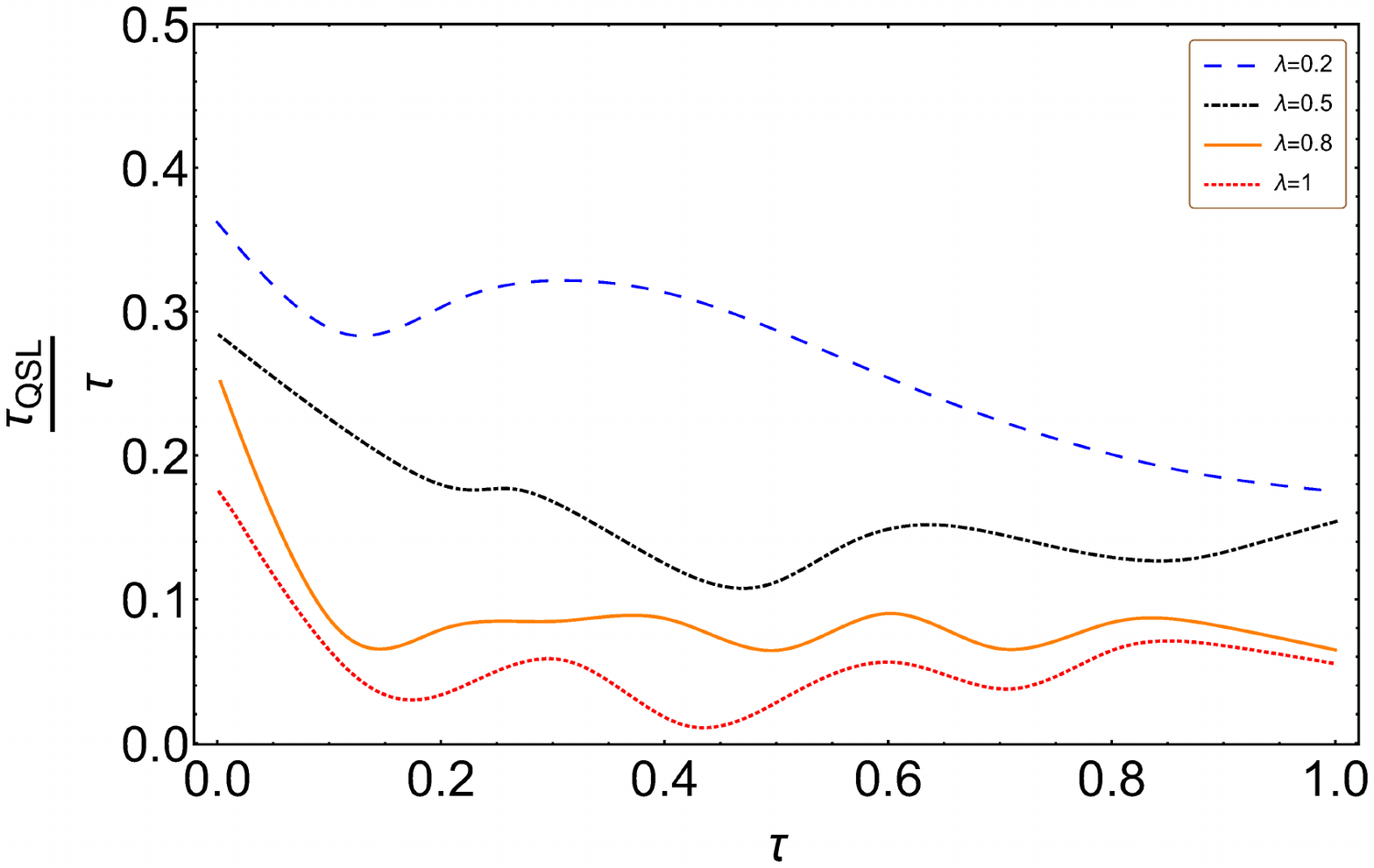}
        \makebox[\linewidth][c]{\small (b2) $\beta=0.35$}
    \end{minipage}

    \begin{minipage}[b]{0.49\linewidth}
        \centering
        \includegraphics[width=\linewidth]{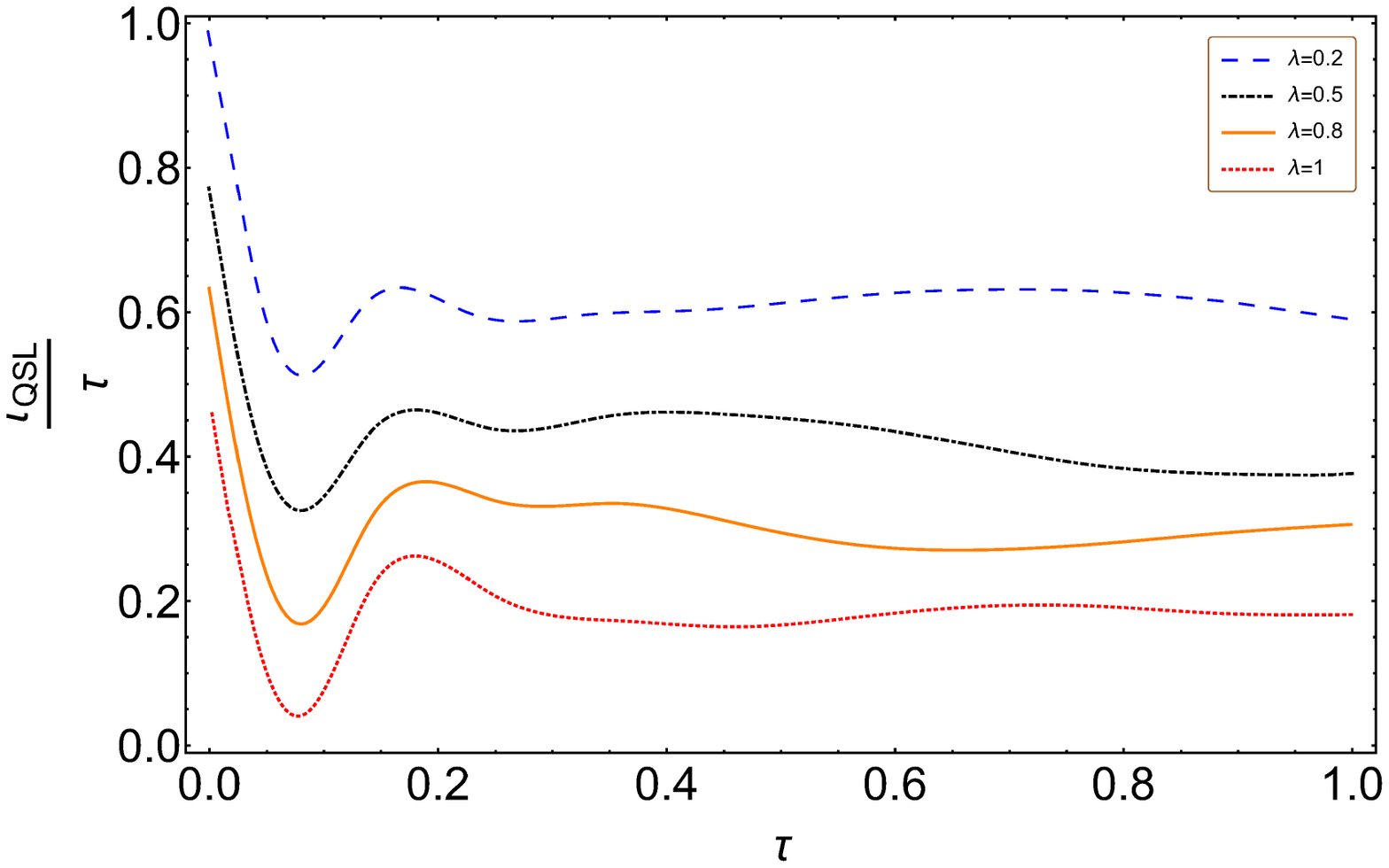}
        \makebox[\linewidth][c]{\small (c1) $\beta=0.65$}
    \end{minipage}
    \hfill
    \begin{minipage}[b]{0.49\linewidth}
        \centering
        \includegraphics[width=\linewidth]{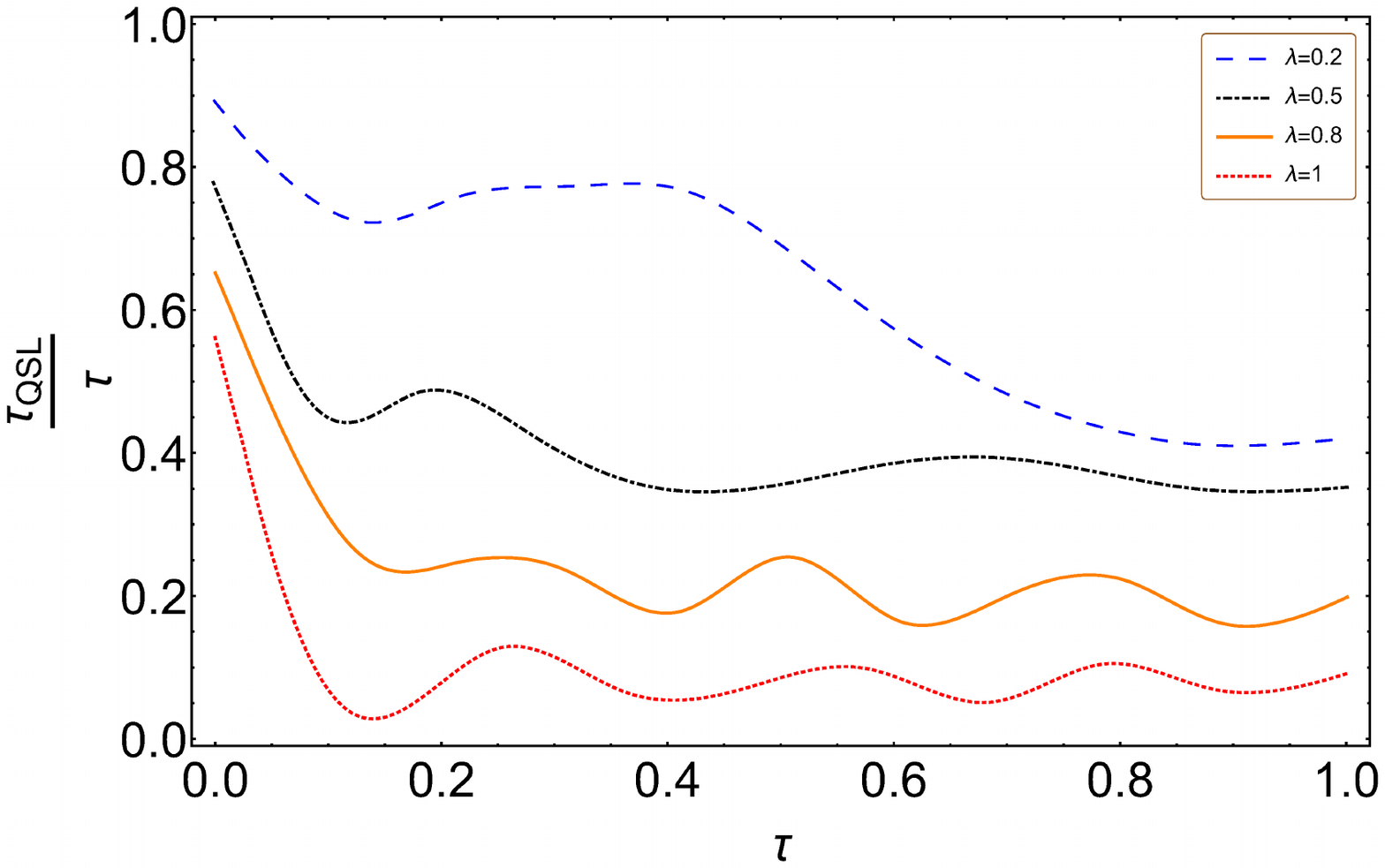}
        \makebox[\linewidth][c]{\small (c2) $\beta=0.65$}
    \end{minipage}

    \begin{minipage}[b]{0.49\linewidth}
        \centering
        \includegraphics[width=\linewidth]{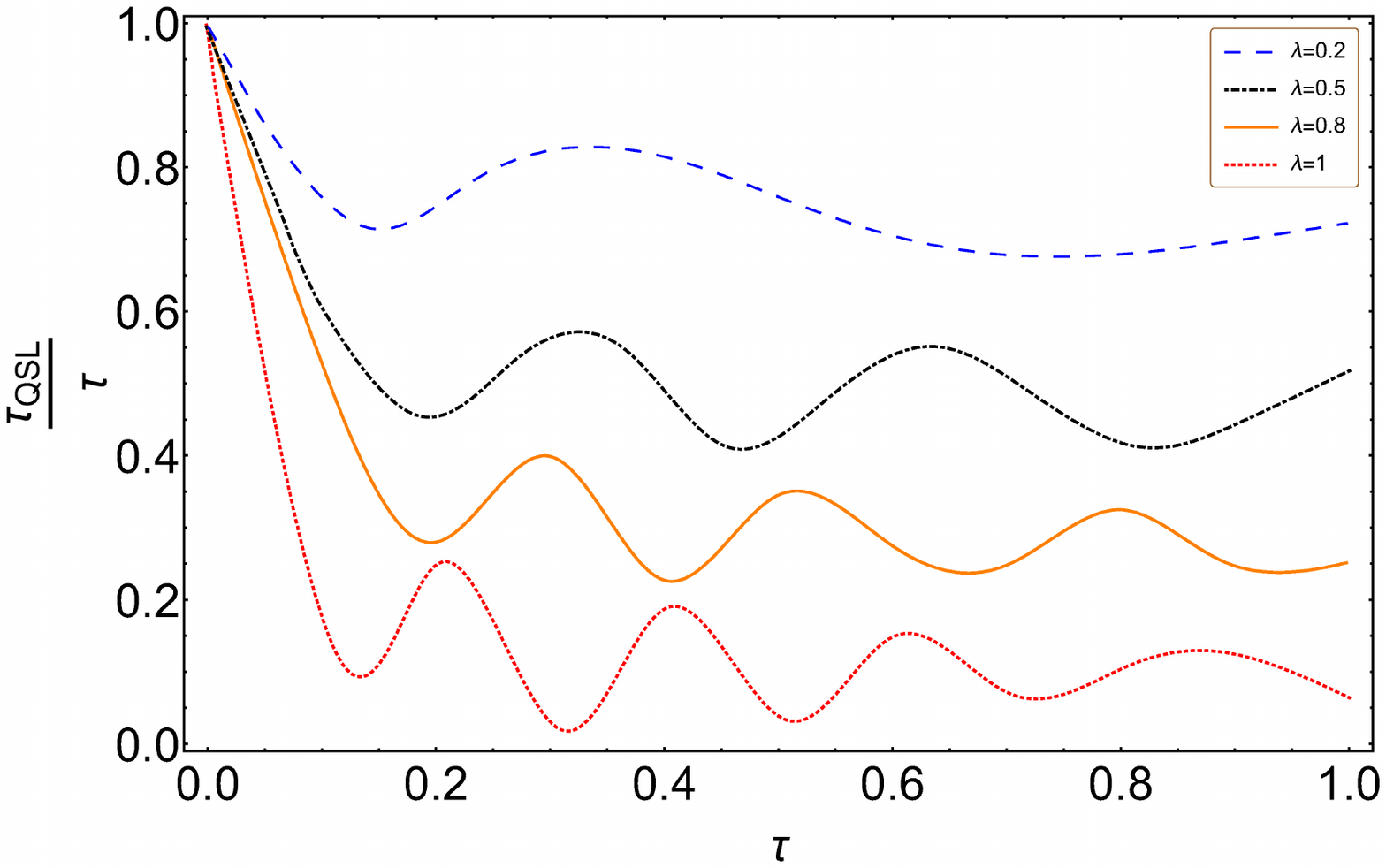}
        \makebox[\linewidth][c]{\small (d1) $\beta=1$}
    \end{minipage}
    \hfill
    \begin{minipage}[b]{0.49\linewidth}
        \centering
        \includegraphics[width=\linewidth]{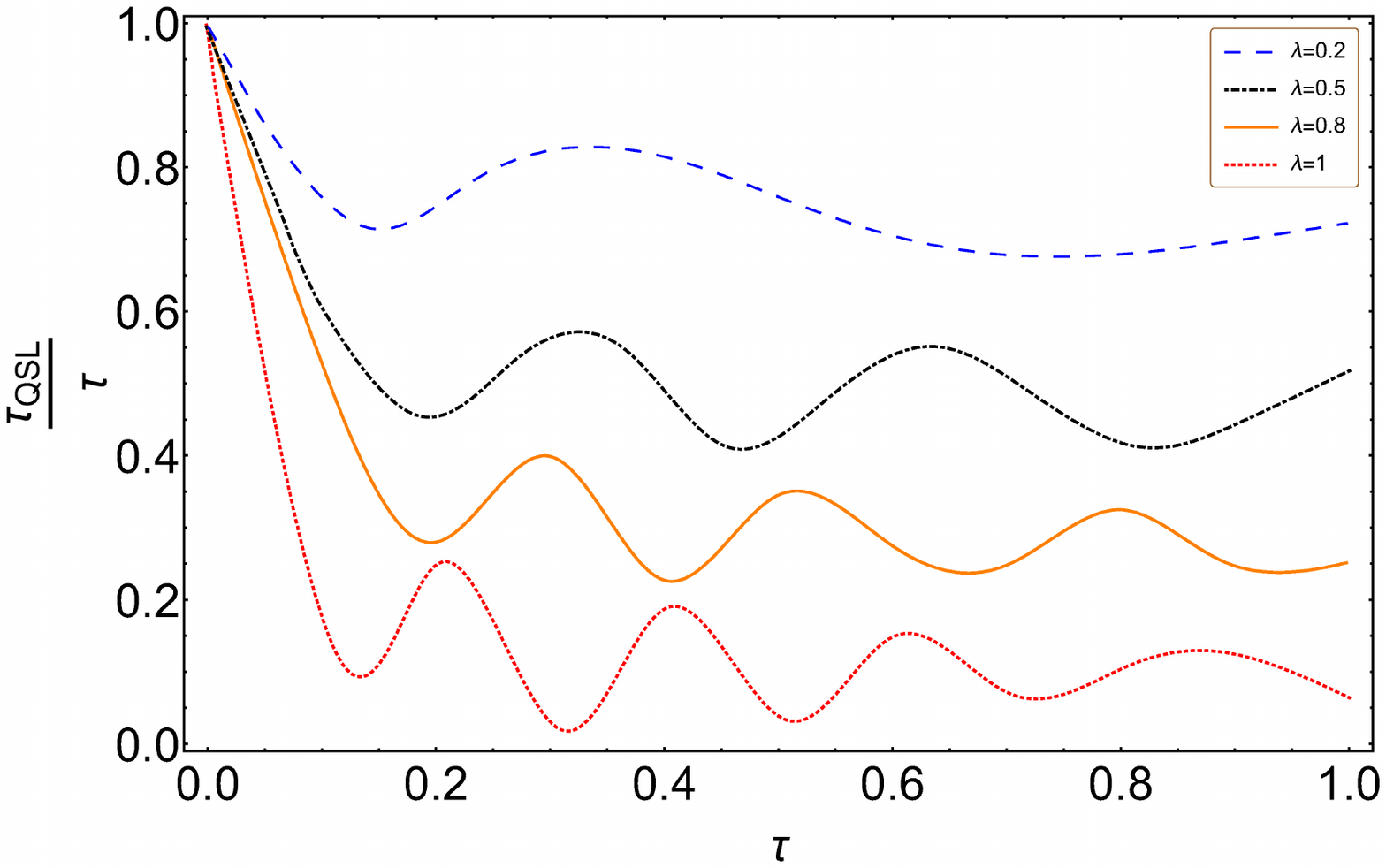}
        \makebox[\linewidth][c]{\small (d2) $\beta=1$}
    \end{minipage}

    \caption{Under the frameworks of Naber's TFSE and Wei's TFSE, the QSL time ratio $\tau_{QSL}/\tau$ of the single-qubit open system as a function of the evolution time $\tau$ for different fractional orders $\beta$ and coupling strengths $\lambda$, where the parameters are set as (a1-a2) $\beta=0.05$, (b1-b2) $\beta=0.35$, (c1-c2) $\beta=0.65$, and (d1-d2) $\beta=1$. Other parameters are $\lambda=0.2,0.5,0.8,1$ and $n=40$.}
    \label{fig:4.3}
\end{figure}

Fig.~\ref{fig:4.3} compares the variation of $\tau_{QSL}/\tau$ with the evolution time $\tau$ under the frameworks of Naber's TFSE and Wei's TFSE for different fractional orders $\beta=0.05,0.35,0.65,1$ and coupling strengths $\lambda=0.2,0.5,0.8,1$, with $n=40$. A detailed analysis is given below.

In Fig.~\ref{fig:4.3}(a1-b1), when $\beta$ takes small or intermediate values, the value of $\tau_{QSL}/\tau$ decreases as $\lambda$ increases, while its amplitude and frequency also show a decreasing trend. However, Fig.~\ref{fig:4.3}(a2-b2) presents another situation. When $\beta$ takes small or intermediate values, the increase in $\lambda$ also leads to a decrease in $\tau_{QSL}/\tau$, but its amplitude and frequency are significantly enhanced, in sharp contrast to the former case. The RDJC model has been verified by both theoretical analysis~\cite{Shenetal2014} and experimental observation~\cite{HuangLiaoKuang2020, Lietal2022}, showing that the non-Markovian oscillatory behavior of the system is enhanced when the coupling between the system and its environment becomes stronger. This conclusion can serve as a criterion for evaluating the performances of TFSE in this model~\cite{Weietal2024}. By comparing Fig.~\ref{fig:4.3}(a1-b1) with Fig.~\ref{fig:4.3}(a2-b2), it can be seen that Wei's TFSE exhibits higher accuracy in describing the accelerated dynamics of the system at small to intermediate $\beta$ values as $\lambda$ gradually increases. Specifically, in Fig.~\ref{fig:4.3}(a1-b1), the increase in $\lambda$ causes the amplitude and frequency of $\tau_{QSL}/\tau$ to weaken simultaneously, which is inconsistent with existing studies. In contrast, in Fig.~\ref{fig:4.3}(a2-b2), Wei's TFSE successfully captures the phenomenon that increasing $\lambda$ simultaneously enhances the amplitude and frequency of $\tau_{QSL}/\tau$, which agrees with existing research expectations. Therefore, Wei's TFSE is more accurate in describing the non-Markovian characteristics of the accelerated dynamics of the system.

In Fig.~\ref{fig:4.3}(c1-d1) and Fig.~\ref{fig:4.3}(c2-d2), when $\beta$ reaches large values, the value of $\tau_{QSL}/\tau$ decreases as $\lambda$ increases, while the oscillation amplitude and frequency show an increasing trend. Notably, in the description of Wei's TFSE, the non-Markovian oscillatory behavior of $\tau_{QSL}/\tau$ is more evident, and its characteristics are more pronounced and clearer than those described by Naber's TFSE. Furthermore, this phenomenon is consistent with the results in Ref.~\cite{DeffnerLutz2013}, which points out that in the dynamics of a single-qubit open system, small to intermediate coupling strengths lead to a plateau feature of the QSL time, whereas the QSL time gradually decreases as $\lambda$ further increases. Similarly, Ref.~\cite{LiuXuZhu2015} also reports that when the environment enters the memory region, namely the strong coupling region, the evolution of a two-qubit open system has greater potential for accelerated evolution, indicating that the QSL time decreases under strong coupling conditions. Therefore, in the descriptions given by both TFSEs, the increase in $\lambda$ induces a decrease in the QSL time, which is consistent with existing literature. In addition, at large $\beta$ values, both TFSEs can successfully capture the simultaneous enhancement of the amplitude and frequency of $\tau_{QSL}/\tau$ caused by the increase in $\lambda$, in agreement with existing research expectations.

\begin{table}[width=\linewidth,cols=3,pos=h]
\centering
\caption{Comparisons of the capability of Naber's TFSE and Wei's TFSE to describe the accelerated dynamics of the system}
\label{tab:3}
\small
\renewcommand{\arraystretch}{1.35}

\begin{tabularx}{\tblwidth}{@{}
  >{\centering\arraybackslash}c
  >{\centering\arraybackslash}X
  >{\centering\arraybackslash}X
@{}}
\toprule
Parameters & Accurate description range of Naber's TFSE & Accurate description range of Wei's TFSE \\
\midrule
Fractional order $\beta$ & $\beta \in (\tilde{\beta}_{QSL},1]$ & $\beta \in (0,1]$ \\
Coupling strength $\lambda$ & $\lambda \in [0,1]$ & $\lambda \in [0,1]$ \\
Photon number $n$ & $n \in \mathbb{N}$ & $n \in \mathbb{N}$ \\
\bottomrule
\end{tabularx}

\smallskip
\begin{minipage}{\tblwidth}
\small
\noindent Note: $\tilde{\beta}_{QSL}\in(0,1]$ denotes the critical point under the framework of Naber's TFSE at which the description of the accelerated evolution process changes from inaccurate to accurate under the influence of the fractional order $\beta$.
\end{minipage}
\end{table}

\subsection{Simulation Efficiency for the Accelerated Dynamics of the System}
\label{sec:Simulation Efficiency for the Accelerated Dynamics of the System}
The simulation of the non-Markovian open quantum system dynamics requires not only that the theoretical tool accurately reflect the memory effects of the environment, but also that the computational process be highly efficient~\cite{DeVegaAlonso2017}. For a time-fractional single-qubit open system, the simulation of its accelerated dynamics is related to $\beta$, $\lambda$, $n$, and $\tau$, making the numerical solution process more complex~\cite{Weietal2023}. Therefore, simulation efficiency has become an important criterion for evaluating the applicability of different TFSE frameworks. Comparing the computational efficiency of Naber's TFSE and Wei's TFSE in simulating the accelerated dynamics helps further clarify their performances differences in the study of the non-Markovian quantum systems.

In terms of the mathematical definition, the TFD adopted in a TFSE affects its computational efficiency. The Ca-FD shown in Eq.~\eqref{eq:1} is defined by differentiating first and then integrating, which is more intuitive and convenient for handling initial value problems. However, because it contains path integration, its computational complexity increases. In contrast, the Co-FD shown in Eq.~\eqref{eq:3} relies only on the basic limit definition of derivatives. It is the simplest, most natural, and most effective definition of a fractional derivative on $\beta\in(0,1]$. Since it does not contain path integration, its computational complexity is lower than that of Ca-FD.

In addition, the form of ACT also determines the computational efficiency of Naber's TFSE and Wei's TFSE. For Naber's TFSE, ACT is given by $t \to t / i\hbar_\beta$. The imaginary unit $i$ is raised to the order of Ca-FD, thereby introducing an $i^\beta$ term and leading to the Mittag-Leffler function $E_\beta(z)$ in the spectral decomposition. For Wei's TFSE, ACT is given by $t \to t / (i\hbar_\beta \beta)^{1/\beta}$, where $i$ is not raised to the order of Co-FD, allowing the equation to retain a purely exponential structure. Therefore, only exponential operations are involved in the spectral decomposition. Since the computational complexity of exponential operations is much lower than that of $E_\beta(z)$, Wei's TFSE has higher computational efficiency in simulating the accelerated dynamics of the system. Table 4 presents the distinctive features of Naber's TFSE~\cite{Naber2004}, Naber's TFSE II~\cite{Naber2004}, XGF's TFSE~\cite{XiangGuoFu2019}, and Wei's TFSE~\cite{Weietal2024}.

\begin{table*}[t]
\centering
\caption{Distinctive features of Naber's TFSE, Naber's TFSE II, XGF's TFSE, and Wei's TFSE}
\label{tab:4}
\small
\renewcommand{\arraystretch}{1.45}
\setlength{\tabcolsep}{4pt}

\begin{tabularx}{\textwidth}{@{}
  >{\centering\arraybackslash}p{0.20\textwidth}
  >{\centering\arraybackslash}p{0.26\textwidth}
  >{\centering\arraybackslash}p{0.24\textwidth}
  >{\centering\arraybackslash}X
@{}}
\toprule
TFSE & TFD & ACT & TFO \\
\midrule
Naber's TFSE~\cite{Naber2004} 
& \shortstack{Ca-FD\\(with path integration)} 
& $t \to t/i\hbar_{\beta}$
& $E_{\beta}\!\left[\left(\dfrac{-it}{\hbar_{\beta}}\right)^{\beta}H_{\beta}\right]$ \\[1.5ex]

Naber's TFSE II~\cite{Naber2004}
& \shortstack{Ca-FD\\(with path integration)} 
& $t \to t/(i\hbar_{\beta}\beta)^{1/\beta}$ 
& $E_{\beta}\!\left(\dfrac{-it^{\beta}}{\hbar_{\beta}}H_{\beta}\right)$ \\[1.5ex]

XGF's TFSE~\cite{XiangGuoFu2019}
& \shortstack{Co-FD\\(without path integration)} 
& $t \to t/i\hbar_{\beta}\beta^{1/\beta}$ 
& $\displaystyle\frac{\left(\dfrac{-i}{\hbar_{\beta}}\right)^{\beta}t^{\beta}H_{\beta}}{e^{\beta}}$ \\[1.5ex]

Wei's TFSE~\cite{Weietal2024}
& \shortstack{Co-FD\\(without path integration)} 
& $t \to t/(i\hbar_{\beta}\beta)^{1/\beta}$ 
& $\displaystyle \frac{\dfrac{-i}{\hbar_{\beta}}t^{\beta}H_{\beta}}{e^{\beta}}$ \\
\bottomrule
\end{tabularx}
\end{table*}

The dynamical trajectories of the excited-state probability $P_e(\tau)$ with respect to $\tau$ under the two TFSE frameworks are characterized separately, and the simulation time is recorded to clarify the simulation efficiency of the two TFSEs.

\begin{figure}[pos=h]
    \centering

    \begin{minipage}[b]{0.49\linewidth}
        \centering
        \includegraphics[width=\linewidth]{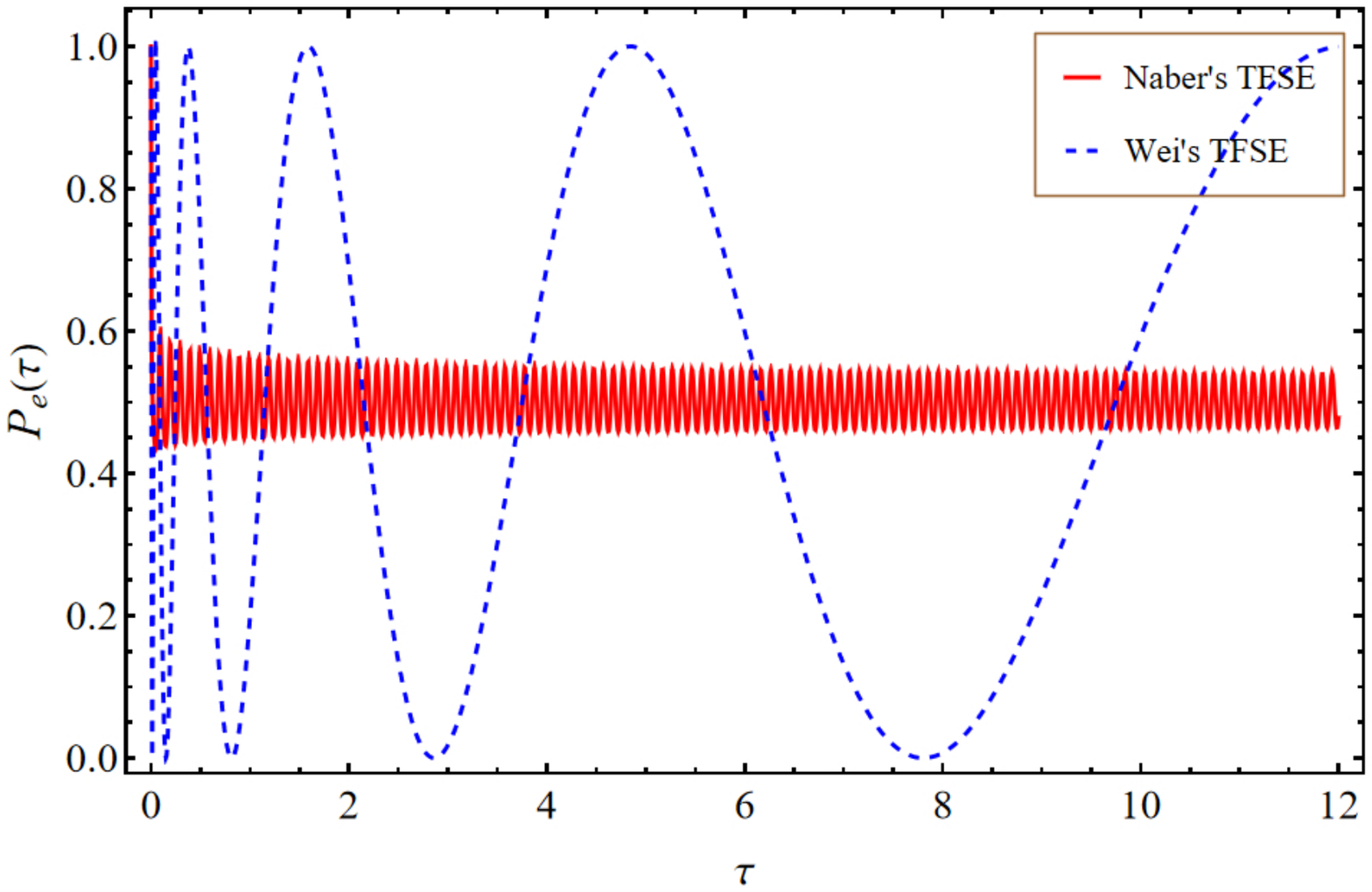}
        \makebox[\linewidth][c]{\small (a) $\beta=0.2$}
    \end{minipage}
    \hfill
    \begin{minipage}[b]{0.49\linewidth}
        \centering
        \includegraphics[width=\linewidth]{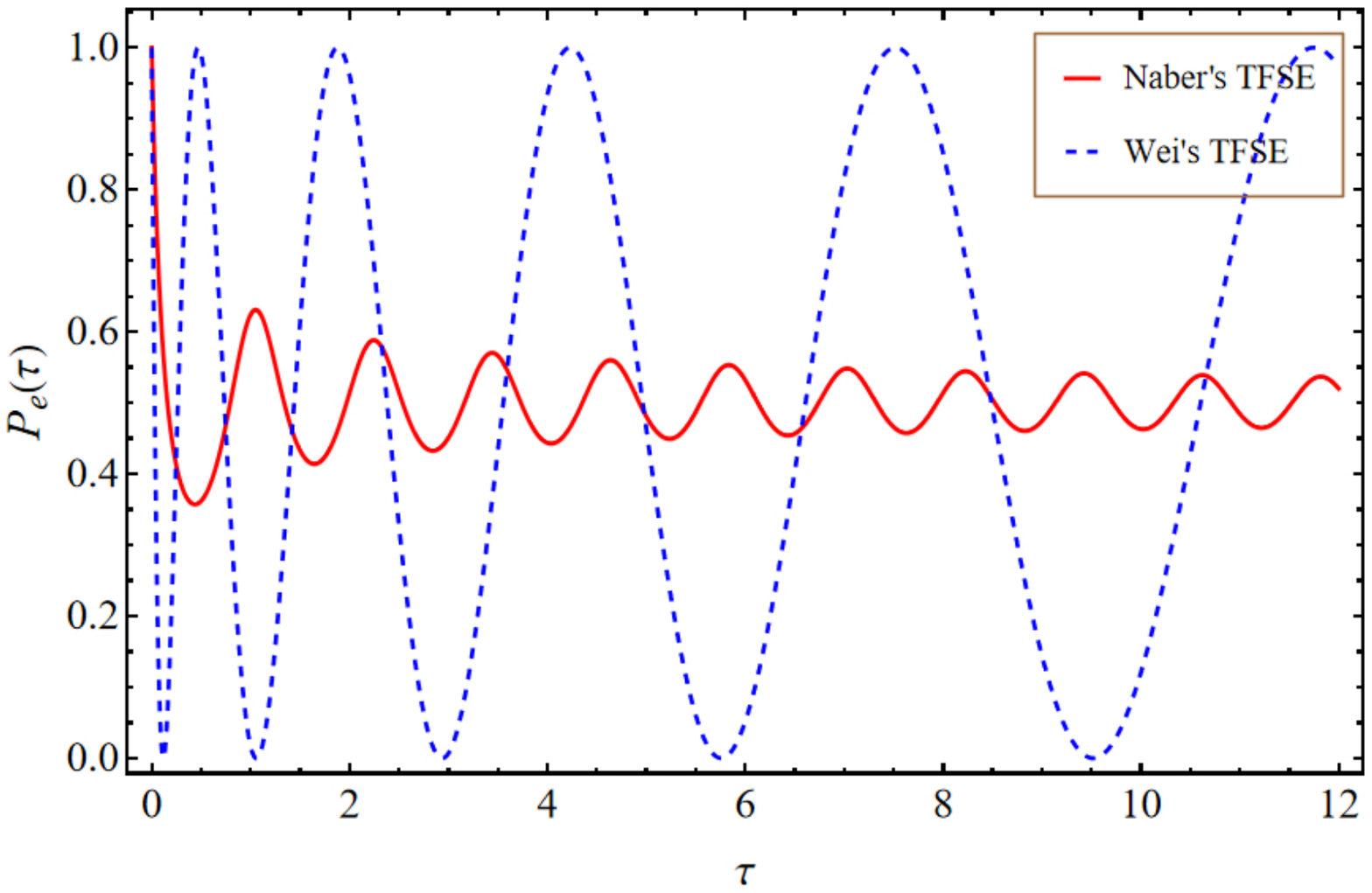}
        \makebox[\linewidth][c]{\small (b) $\beta=0.5$}
    \end{minipage}

    \begin{minipage}[b]{0.49\linewidth}
        \centering
        \includegraphics[width=\linewidth]{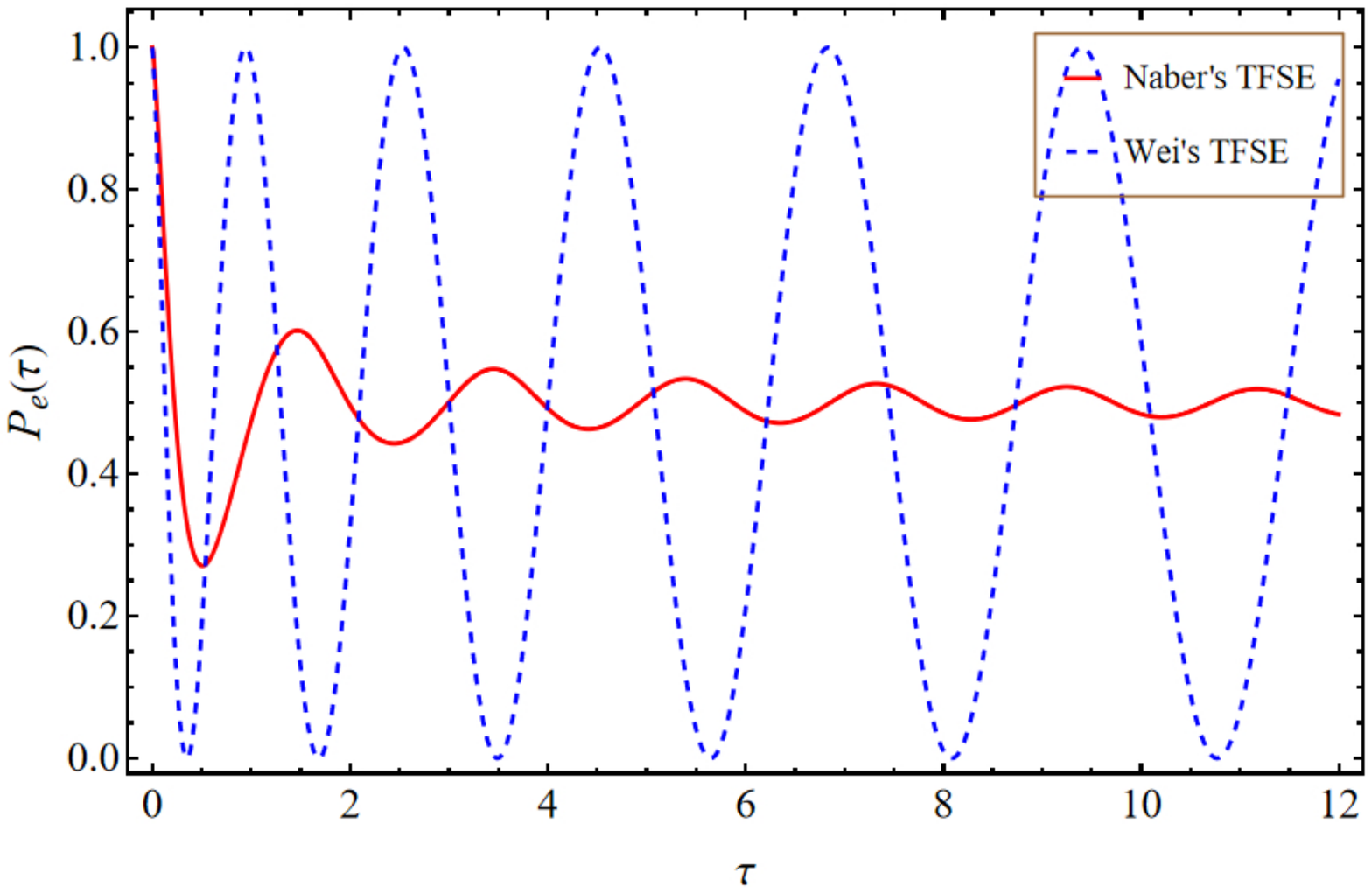}
        \makebox[\linewidth][c]{\small (c) $\beta=0.7$}
    \end{minipage}
    \hfill
    \begin{minipage}[b]{0.49\linewidth}
        \centering
        \includegraphics[width=\linewidth]{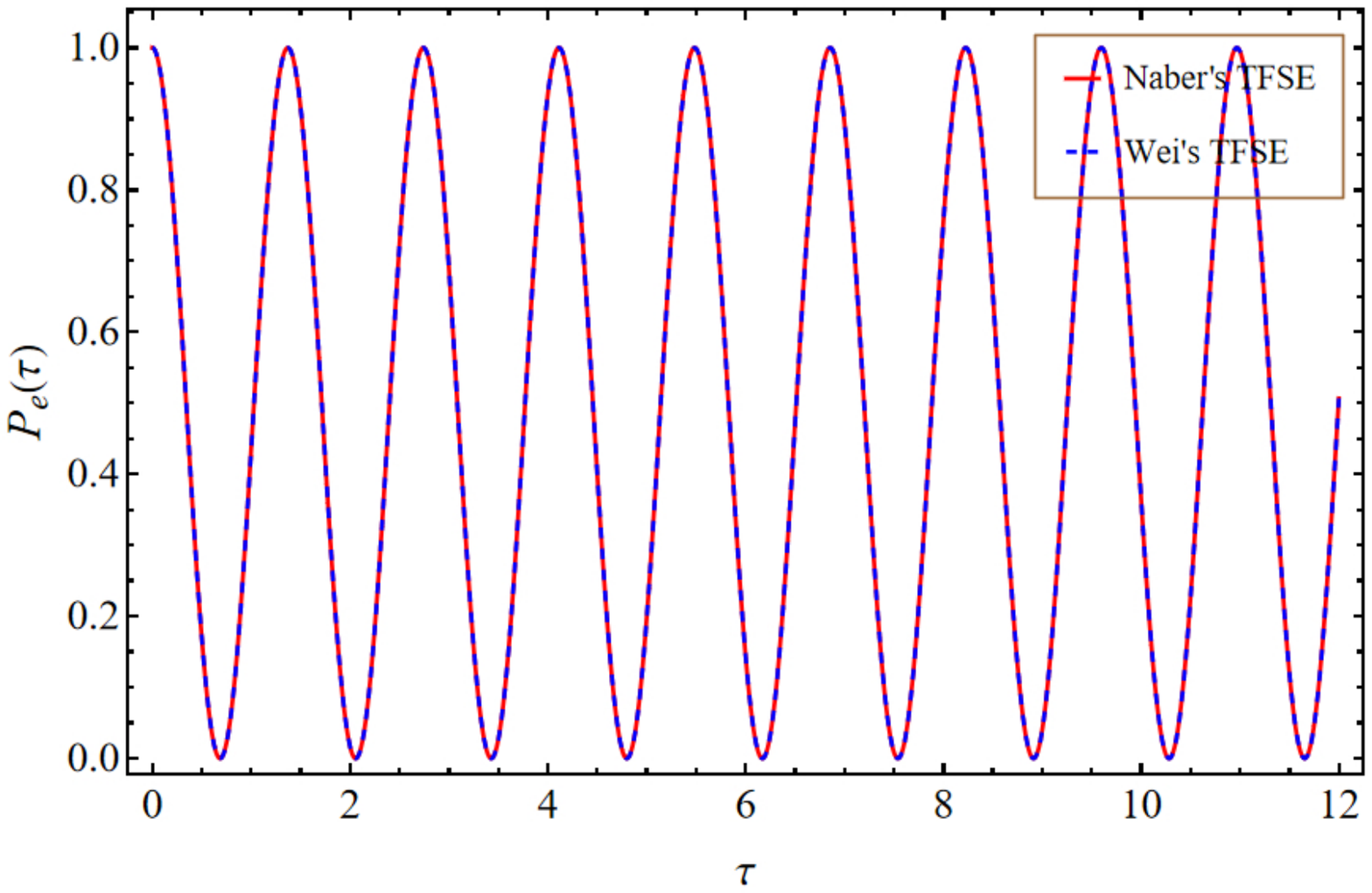}
        \makebox[\linewidth][c]{\small (d) $\beta=1$}
    \end{minipage}

    \caption{Dynamical trajectories of the excited-state probability $P_e(\tau)$ as a function of the evolution time $\tau$ under Naber's TFSE and Wei's TFSE for different fractional orders $\beta=0.2,0.5,0.7,1$. The coupling strength is set as $\lambda=0.5$, the photon number as $n=20$, and the number of sampling points as $2000$.}
    \label{fig:4.4}
\end{figure}

Fig.~\ref{fig:4.4} shows the evolution results of the excited-state probability $P_e(\tau)$ with respect to $\tau$ under the two TFSE frameworks. The time evolution of Naber's TFSE is controlled by $E_\beta(z)$, so the oscillation amplitude of $P_e(\tau)$ is significantly suppressed, mainly showing a rapid decay followed by a slow recovery and then a plateau. This indicates that Naber's TFSE has insufficient capability in characterizing the non-Markovian memory effects. In contrast, $P_e(\tau)$ under Wei's TFSE exhibits significant and persistent non-Markovian oscillations. This shows that Wei's TFSE can more accurately capture the phenomenon of information backflow between the system and the environment. When $\beta=1$, both TFSEs reduce to the integer-order Schr\"odinger equation, and the resulting dynamical trajectories are identical.

It can be seen from Table~\ref{tab:simulation_time_comparison} that the average simulation time of Wei's TFSE is significantly lower than that of Naber's TFSE. The computational time of Naber's TFSE increases non-monotonically with $\beta$. This is because the main computational overhead of Naber's TFSE comes from $E_\beta(z)$, whose numerical complexity depends not only on $\beta$, but also on the value range of the complex variable \mbox{$z=(-i)^\beta \tau^\beta \lambda\sqrt{n+1}$}. When $\beta=0.2$, $\tau^\beta$ grows slowly over the given time interval $\tau\in[0,12]$, making the modulus of $z$ in $E_\beta(z)$ relatively small; thus the computational time is relatively low. When $\beta=0.5$ and $\beta=0.7$, the value range of $z$ expands, increasing the difficulty of numerical computation and leading to a marked increase in the required simulation time. In particular, when $\beta=1$, $E_\beta(z)$ degenerates into the ordinary exponential function, so the computational complexity is greatly reduced and the simulation time of Naber's TFSE is significantly shortened. By comparison, the computation of Wei's TFSE does not require $E_\beta(z)$, and its computational time remains essentially stable under different $\beta$ values. These results show that Wei's TFSE has advantages in both characterizing non-Markovian features and improving computational efficiency when simulating non-Markovian quantum accelerated dynamics.

\begin{table}[width=\linewidth,cols=4,pos=h]
\centering
\caption{Comparisons of the average simulation time required to calculate the dynamical trajectory of the excited-state probability $P_e(\tau)$ using Naber's TFSE and Wei's TFSE}
\label{tab:simulation_time_comparison}
\small
\setlength{\tabcolsep}{3pt}
\renewcommand{\arraystretch}{1.25}

\begin{tabularx}{\tblwidth}{@{}
  >{\centering\arraybackslash}c
  >{\centering\arraybackslash}X
  >{\centering\arraybackslash}X
  >{\centering\arraybackslash}X
@{}}
\toprule
Fractional order $\beta$ 
& Average simulation time of Naber's TFSE / s 
& Average simulation time of Wei's TFSE / s 
& Average simulation time ratio of the two TFSEs \\
\midrule
$\beta = 0.2$ & 3.58  & 0.05 & 66.53  \\
$\beta = 0.5$ & 13.01 & 0.05 & 256.29 \\
$\beta = 0.7$ & 16.37 & 0.05 & 269.90 \\
$\beta = 1$   & 0.42  & 0.05 & 8.28   \\
\bottomrule
\end{tabularx}

\begin{minipage}{\tblwidth}
\small
\noindent Note: The simulation time in the table is obtained by generating $2000$ sampling points over the evolution time interval \mbox{$\tau\in[0,12]$} and averaging the results over $7$ repeated simulations.
\end{minipage}
\end{table}

\section{Conclusion}\label{sec:Conclusion}
By analyzing the QSL time and the dynamical trajectory of the excited-state probability, this paper compares the performances of Naber's TFSE and Wei's TFSE in simulating the non-Markovian quantum accelerated dynamics. The results show that Wei's TFSE is more accurate and efficient in simulating the accelerated dynamics of the time-fractional open quantum system. Specifically, the enhancement mechanism of the system evolution speed induced by the non-Markovian memory effects of the environment is revealed. Furthermore, the optimized acceleration of the system evolution can be achieved by jointly regulating the fractional order, coupling strength, and photon number. These findings are consistent with those obtained under the framework of Naber's TFSE. However, Wei's TFSE performs better than Naber's TFSE in describing the accelerated evolution process of the system. In the RDJC model, when the system interacts with the external environment, Wei's TFSE can accurately capture the non-Markovian accelerated dynamical features of the system over the entire fractional order range $\beta\in(0,1]$, whereas Naber's TFSE can do so only for large values of $\beta$. In addition, Wei's TFSE based on Co-FD has a significant simulation advantage in computational efficiency over Naber's TFSE based on Ca-FD. In future work, Wei's TFSE can be further combined with methods such as the process tensor~\cite{Cygoreketal2022} and memory kernel construction~\cite{IvanderLindoyLee2024} to establish a scalable simulation framework for non-Markovian dynamics.


\printcredits

\bibliographystyle{unsrtnat}

\bibliography{cas-refs}

@BOOK{BreuerPetruccione2002,
  author    = {Breuer, H. P. and Petruccione, F.},
  title     = {The Theory of Open Quantum Systems},
  publisher = {Oxford University Press},
  address   = {Oxford},
  year      = {2002},
  doi       = {10.1093/acprof:oso/9780199213900.001.0001}
}

@ARTICLE{Lindblad1976,
  author  = {Lindblad, G.},
  title   = {On the generators of quantum dynamical semigroups},
  journal = {Commun. Math. Phys.},
  volume  = {48},
  number  = {2},
  year    = {1976},
  pages   = {119-130},
  doi     = {10.1007/BF01608499}
}

@ARTICLE{Breueretal2016,
  author  = {Breuer, H. P. and Laine, E. M. and Piilo, J. and others},
  title   = {Colloquium: {Non-Markovian} dynamics in open quantum systems},
  journal = {Rev. Mod. Phys.},
  volume  = {88},
  number  = {2},
  year    = {2016},
  pages   = {021002},
  doi     = {10.1103/RevModPhys.88.021002}
}

@ARTICLE{Zurek2003,
  author  = {Zurek, W. H.},
  title   = {Decoherence, einselection, and the quantum origins of the classical},
  journal = {Rev. Mod. Phys.},
  volume  = {75},
  number  = {3},
  year    = {2003},
  pages   = {715},
  doi     = {10.1103/RevModPhys.75.715}
}

@BOOK{Weiss2012,
  author    = {Weiss, U.},
  title     = {Quantum Dissipative Systems},
  publisher = {World Scientific},
  year      = {2012},
  doi       = {10.1142/8334}
}

@ARTICLE{BreuerLainePiilo2009,
  author  = {Breuer, H. P. and Laine, E. M. and Piilo, J.},
  title   = {Measure for the degree of {Non-Markovian} behavior of quantum processes in open systems},
  journal = {Phys. Rev. Lett.},
  volume  = {103},
  number  = {21},
  year    = {2009},
  pages   = {210401},
  doi     = {10.1103/PhysRevLett.103.210401}
}

@ARTICLE{Davies1974,
  author  = {Davies, E. B.},
  title   = {Markovian master equations},
  journal = {Commun. Math. Phys.},
  volume  = {39},
  number  = {2},
  year    = {1974},
  pages   = {91-110},
  doi     = {10.1007/BF01608389}
}

@ARTICLE{Alickietal2004,
  author  = {Alicki, R. and Horodecki, M. and Horodecki, P. and others},
  title   = {Optimal strategy for a single-qubit gate and the trade-off between opposite types of decoherence},
  journal = {Phys. Rev. A},
  volume  = {70},
  number  = {1},
  year    = {2004},
  pages   = {010501(R)},
  doi     = {10.1103/PhysRevA.70.010501}
}

@ARTICLE{Daffereretal2004,
  author  = {Daffer, S. and W{\'o}dkiewicz, K. and Cresser, J. D. and others},
  title   = {Depolarizing channel as a completely positive map with memory},
  journal = {Phys. Rev. A},
  volume  = {70},
  number  = {1},
  year    = {2004},
  pages   = {010304(R)},
  doi     = {10.1103/PhysRevA.70.010304}
}

@ARTICLE{TerhalBurkard2005,
  author  = {Terhal, B. M. and Burkard, G.},
  title   = {Fault-tolerant quantum computation for local {Non-Markovian} noise},
  journal = {Phys. Rev. A},
  volume  = {71},
  number  = {1},
  year    = {2005},
  pages   = {012336},
  doi     = {10.1103/PhysRevA.71.012336}
}

@ARTICLE{AliferisGottesmanPreskill2006,
  author  = {Aliferis, P. and Gottesman, D. and Preskill, J.},
  title   = {Quantum accuracy threshold for concatenated distance-3 codes},
  journal = {Quantum Inf. Comput.},
  volume  = {6},
  year    = {2006},
  pages   = {97-165},
  doi     = {10.48550/arXiv.quant-ph/0504218}
}

@ARTICLE{Carusoetal2014,
  author  = {Caruso, F. and Giovannetti, V. and Lupo, C. and others},
  title   = {Quantum channels and memory effects},
  journal = {Rev. Mod. Phys.},
  volume  = {86},
  number  = {4},
  year    = {2014},
  pages   = {1203},
  doi     = {10.1103/RevModPhys.86.1203}
}

@ARTICLE{Meleetal2024,
  author  = {Mele, F. A. and De Palma, G. and Fanizza, M. and others},
  title   = {Optical fibers with memory effects and their quantum communication capacities},
  journal = {IEEE Trans. Inf. Theory},
  volume  = {70},
  number  = {12},
  year    = {2024},
  pages   = {8844-8869},
  doi     = {10.1109/TIT.2024.3450501}
}

@ARTICLE{Miaoetal2025,
  author  = {Miao, R. H. and Liu, Z. D. and Ning, C. X. and others},
  title   = {Implementation of multiparticle quantum speed limits on observables},
  journal = {Sci. Adv.},
  volume  = {11},
  number  = {44},
  year    = {2025},
  pages   = {eadk8765},
  doi     = {10.1126/sciadv.ady0497}
}

@ARTICLE{Gaikwadetal2024,
  author  = {Gaikwad, C. and Kowsari, D. and Brame, C. and others},
  title   = {Entanglement assisted probe of the {Non-Markovian} to {Markovian} transition in open quantum system dynamics},
  journal = {Phys. Rev. Lett.},
  volume  = {132},
  number  = {20},
  year    = {2024},
  pages   = {200401},
  doi     = {10.1103/PhysRevLett.132.200401}
}

@ARTICLE{Lorenzonietal2025,
  author  = {Lorenzoni, N. and Lacroix, T. and Lim, J. and others},
  title   = {Full microscopic simulations uncover persistent quantum effects in primary photosynthesis},
  journal = {Sci. Adv.},
  volume  = {11},
  number  = {40},
  year    = {2025},
  pages   = {eady6751},
  doi     = {10.1126/sciadv.ady6751}
}

@ARTICLE{Iomin2009,
  author  = {Iomin, A.},
  title   = {Fractional-time quantum dynamics},
  journal = {Phys. Rev. E},
  volume  = {80},
  number  = {2},
  year    = {2009},
  pages   = {022103},
  doi     = {10.1103/PhysRevE.80.022103}
}

@ARTICLE{Wuetal2010,
  author  = {Wu, J. N. and Huang, C. H. and Cheng, S. C. and others},
  title   = {Spontaneous emission from a two-level atom in anisotropic one-band photonic crystals: A fractional calculus approach},
  journal = {Phys. Rev. A},
  volume  = {81},
  number  = {2},
  year    = {2010},
  pages   = {023827},
  doi     = {10.1103/PhysRevA.81.023827}
}

@ARTICLE{Huangetal2011,
  author  = {Huang, C. H. and Wu, J. N. and Li, Y. Y. and others},
  title   = {Calculation of spontaneous emission from a {V}-type three-level atom in photonic crystals using fractional calculus},
  journal = {Phys. Rev. A},
  volume  = {84},
  number  = {1},
  year    = {2011},
  pages   = {013802},
  doi     = {10.1103/PhysRevA.84.013802}
}

@ARTICLE{ZhaoLuo2019,
  author  = {Zhao, D. Z. and Luo, M. K.},
  title   = {Representations of acting processes and memory effects: General fractional derivative and its application to theory of heat conduction with finite wave speeds},
  journal = {Appl. Math. Comput.},
  volume  = {346},
  year    = {2019},
  pages   = {531-544},
  doi     = {10.1016/j.amc.2018.10.037}
}

@ARTICLE{Ertiketal2010,
  author  = {Ertik, H. and Demirhan, D. and {\c S}irin, H. and others},
  title   = {Time fractional development of quantum systems},
  journal = {J. Math. Phys.},
  volume  = {51},
  number  = {8},
  year    = {2010},
  pages   = {082102},
  doi     = {10.1063/1.3464492}
}

@ARTICLE{Laskin2017,
  author  = {Laskin, N.},
  title   = {Time fractional quantum mechanics},
  journal = {Chaos Solitons Fractals},
  volume  = {102},
  year    = {2017},
  pages   = {16-28},
  doi     = {10.1016/j.chaos.2017.04.010}
}

@ARTICLE{Naber2004,
  author  = {Naber, M.},
  title   = {Time fractional {Schr{\"o}dinger} equation},
  journal = {J. Math. Phys.},
  volume  = {45},
  number  = {8},
  year    = {2004},
  pages   = {3339},
  doi     = {10.1063/1.1769611}
}

@ARTICLE{Mainardi1996,
  author  = {Mainardi, F.},
  title   = {Fractional relaxation-oscillation and fractional diffusion-wave phenomena},
  journal = {Chaos Solitons Fractals},
  volume  = {7},
  number  = {9},
  year    = {1996},
  pages   = {1461-1477},
  doi     = {10.1016/0960-0779(95)00125-5}
}

@ARTICLE{ZuYu2022,
  author  = {Zu, C. and Yu, X.},
  title   = {Time fractional {Schr{\"o}dinger} equation with a limit based fractional derivative},
  journal = {Chaos Solitons Fractals},
  volume  = {157},
  year    = {2022},
  pages   = {111941},
  doi     = {10.1016/j.chaos.2022.111941}
}

@ARTICLE{WangXu2007,
  author  = {Wang, S. and Xu, M. Y.},
  title   = {Generalized fractional {Schr{\"o}dinger} equation with space-time fractional derivatives},
  journal = {J. Math. Phys.},
  volume  = {48},
  number  = {4},
  year    = {2007},
  pages   = {043502},
  doi     = {10.1063/1.2716203}
}

@ARTICLE{DongXu2008,
  author  = {Dong, J. and Xu, M.},
  title   = {Space-time fractional {Schr{\"o}dinger} equation with time-independent potentials},
  journal = {J. Math. Anal. Appl.},
  volume  = {344},
  number  = {2},
  year    = {2008},
  pages   = {1005-1017},
  doi     = {10.1016/j.jmaa.2008.03.061}
}

@ARTICLE{XiangGuoFu2019,
  author  = {Xiang, P. and Guo, Y. X. and Fu, J. L.},
  title   = {Time and space fractional {Schr{\"o}dinger} equation with fractional factor},
  journal = {Commun. Theor. Phys.},
  volume  = {71},
  year    = {2019},
  pages   = {16-26},
  doi     = {10.1088/0253-6102/71/1/16}
}

@ARTICLE{Weietal2024,
  author  = {Wei, D. M. and Liu, H. L. and Li, Y. M. and others},
  title   = {{Non-Markovian} dynamics of time-fractional open quantum systems},
  journal = {Chaos Solitons Fractals},
  volume  = {182},
  year    = {2024},
  pages   = {114816},
  doi     = {10.1016/j.chaos.2024.114816}
}

@ARTICLE{DeffnerCampbell2017,
  author  = {Deffner, S. and Campbell, S.},
  title   = {Quantum speed limits: From {Heisenberg}'s uncertainty principle to optimal quantum control},
  journal = {J. Phys. A: Math. Theor.},
  volume  = {50},
  number  = {45},
  year    = {2017},
  pages   = {453001},
  doi     = {10.1088/1751-8121/aa86c6}
}

@ARTICLE{Taddeietal2013,
  author  = {Taddei, M. M. and Escher, B. M. and Davidovich, L. and others},
  title   = {Quantum speed limit for physical processes},
  journal = {Phys. Rev. Lett.},
  volume  = {110},
  number  = {5},
  year    = {2013},
  pages   = {050402},
  doi     = {10.1103/PhysRevLett.110.050402}
}

@ARTICLE{Lloyd2000,
  author  = {Lloyd, S.},
  title   = {Ultimate physical limits to computation},
  journal = {Nature},
  volume  = {406},
  number  = {6799},
  year    = {2000},
  pages   = {1047-1054},
  doi     = {10.1038/35023282}
}

@ARTICLE{HerbDegen2024,
  author  = {Herb, K. and Degen, C. L.},
  title   = {Quantum speed limit in quantum sensing},
  journal = {Phys. Rev. Lett.},
  volume  = {133},
  number  = {21},
  year    = {2024},
  pages   = {210802},
  doi     = {10.1103/PhysRevLett.133.210802}
}

@ARTICLE{Zhuetal2025,
  author  = {Zhu, Z. and Gao, L. and Bao, Z. and others},
  title   = {Observation of minimal and maximal speed limits for few and many-body states},
  journal = {Nat. Commun.},
  volume  = {16},
  number  = {1},
  year    = {2025},
  pages   = {1255},
  doi     = {10.1038/s41467-025-56451-3}
}

@ARTICLE{Weietal2023,
  author  = {Wei, D. M. and Liu, H. L. and Li, Y. M. and others},
  title   = {Quantum speed limit for time-fractional open systems},
  journal = {Chaos Solitons Fractals},
  volume  = {175},
  year    = {2023},
  pages   = {114065},
  doi     = {10.1016/j.chaos.2023.114065}
}

@ARTICLE{DeffnerLutz2013,
  author  = {Deffner, S. and Lutz, E.},
  title   = {Quantum speed limit for {Non-Markovian} dynamics},
  journal = {Phys. Rev. Lett.},
  volume  = {111},
  number  = {1},
  year    = {2013},
  pages   = {010402},
  doi     = {10.1103/PhysRevLett.111.010402}
}

@ARTICLE{Shenetal2014,
  author  = {Shen, H. Z. and Qin, M. and Xiu, X. M. and others},
  title   = {Exact {Non-Markovian} master equation for a driven damped two-level system},
  journal = {Phys. Rev. A},
  volume  = {89},
  number  = {6},
  year    = {2014},
  pages   = {062113},
  doi     = {10.1103/PhysRevA.89.062113}
}

@ARTICLE{HuangLiaoKuang2020,
  author  = {Huang, J. F. and Liao, J. Q. and Kuang, L. M.},
  title   = {Ultrastrong {Jaynes-Cummings} model},
  journal = {Phys. Rev. A},
  volume  = {101},
  number  = {4},
  year    = {2020},
  pages   = {043835},
  doi     = {10.1103/PhysRevA.101.043835}
}

@ARTICLE{Lietal2022,
  author  = {Li, B. W. and Mei, Q. X. and Wu, Y. K. and others},
  title   = {Observation of {Non-Markovian} spin dynamics in a {Jaynes-Cummings-Hubbard} model using a trapped-ion quantum simulator},
  journal = {Phys. Rev. Lett.},
  volume  = {129},
  number  = {14},
  year    = {2022},
  pages   = {140501},
  doi     = {10.1103/PhysRevLett.129.140501}
}

@ARTICLE{LiuXuZhu2015,
  author  = {Liu, C. and Xu, Z. Y. and Zhu, S. Q.},
  title   = {Quantum-speed-limit time for multiqubit open systems},
  journal = {Phys. Rev. A},
  volume  = {91},
  number  = {2},
  year    = {2015},
  pages   = {022102},
  doi     = {10.1103/PhysRevA.91.022102}
}

@ARTICLE{DeVegaAlonso2017,
  author  = {De Vega, I. and Alonso, D.},
  title   = {Dynamics of {Non-Markovian} open quantum systems},
  journal = {Rev. Mod. Phys.},
  volume  = {89},
  number  = {1},
  year    = {2017},
  pages   = {015001},
  doi     = {10.1103/RevModPhys.89.015001}
}

@ARTICLE{Cygoreketal2022,
  author  = {Cygorek, M. and Cosacchi, M. and Vagov, A. and others},
  title   = {Simulation of open quantum systems by automated compression of arbitrary environments},
  journal = {Nat. Phys.},
  volume  = {18},
  year    = {2022},
  pages   = {662-668},
  doi     = {10.1038/s41567-022-01544-9}
}

@ARTICLE{IvanderLindoyLee2024,
  author  = {Ivander, F. and Lindoy, L. P. and Lee, J.},
  title   = {Unified framework for open quantum dynamics with memory},
  journal = {Nat. Commun.},
  volume  = {15},
  year    = {2024},
  pages   = {8087},
  doi     = {10.1038/s41467-024-52081-3}
}



\end{document}